\documentclass[12pt,letterpaper]{article}
\usepackage{html}
\usepackage[dvipdf]{graphics}
\usepackage[all,arc,dvips]{xy}
\usepackage{pstricks}
\usepackage{pst-plot}
\renewcommand{\thesection}{\Roman{section}.}
\renewcommand{\baselinestretch}{1.5}
\title{EXPERIMENTAL STUDY\\
 OF THE LASER DIODE PUMPED\\
  RUBIDIUM MASER
\thanks{{\em IEEE Trans. Instrum. Meas.}, vol. 40, no 2, pp. 170-3, April 1991. This work has been presented at the CPEM'90, Ottawa, Canada, June 1990, DOI:10.1109/CPEM.1990.109969, 	 Available: [online] .}} 
\author{Alain Michaud, Pierre Tremblay \\
   and Michel T\^{e}tu                 \\
{\em Centre d'optique, photonique et laser (COPL),}		\\
{\em D\'{e}p. de g\'{e}nie \'{e}lectrique, Universit\'{e} Laval,}\\
{\em Qu\'{e}bec (Qu\'{e}bec) G1K 7P4, Canada}			\\}
\date{}
\begin{document}
\maketitle
%
%
\begin{abstract}
We report the operation of a $^{87}$Rb maser in the
self-oscillating mode, using a laser diode as
the optical pumping source. The maser uses a TE$_{021}$ cavity surrounding a
cell containing the atoms and 11 Torr of N$_{2}$ as the buffer gas.
The optical
pumping is accomplished by using a commercial  laser diode frequency-locked
on the linear absorption line of an external rubidium cell.
The maser output power is maximized when the laser spectrum
is spread by modulating
its frequency through the variation of its injection current.
The maser output power is also presented for various
modulation waveforms and frequencies. We find that the spurious sidebands
induced by the modulation can be minimized. Finally, the linear dependence
between the laser and maser frequencies is shown.
We find that the relative maser frequency pulling from the laser is
about $4.6 \times 10^{-3}$.

\end{abstract}

\section{Introduction}

The availability of high quality, low cost laser diodes, brings
many opportunities in metrology. This is especially true in optical pumping
experiments where their tunability permits the replacement of the traditional
spectral lamp or the cumbersome dye laser. Recently, many papers reported their use for the optical
pumping of various alkalis such as cesium \cite{DEC84}, \cite{OHS88}. We used a
similar technique  to optically pump a rubidium maser, replacing the usual
lamp-isotopic filter configuration by a laser diode.
Two reasons motivate this study.
The conventional
active frequency standard showed
a frequency stability of  about $5 \times 10^{-14}$ for averaging times
of the order of 100 s, but starts to deteriorate for
longer times \cite{TET85}. The light fluctuations
were supposed to be a major cause for the long-term instabilities.
Optical pumping from the laser brings the possibility of a better control
of these fluctuations.
Also proper operation of the maser involves three
temperatures for the cavity, the filter and the lamp.
The gradients of temperature present, and the volume of the elements
complicate the design of a compact, temperature stabilized
optical pumping module.

\section{Description}

\begin{figure}
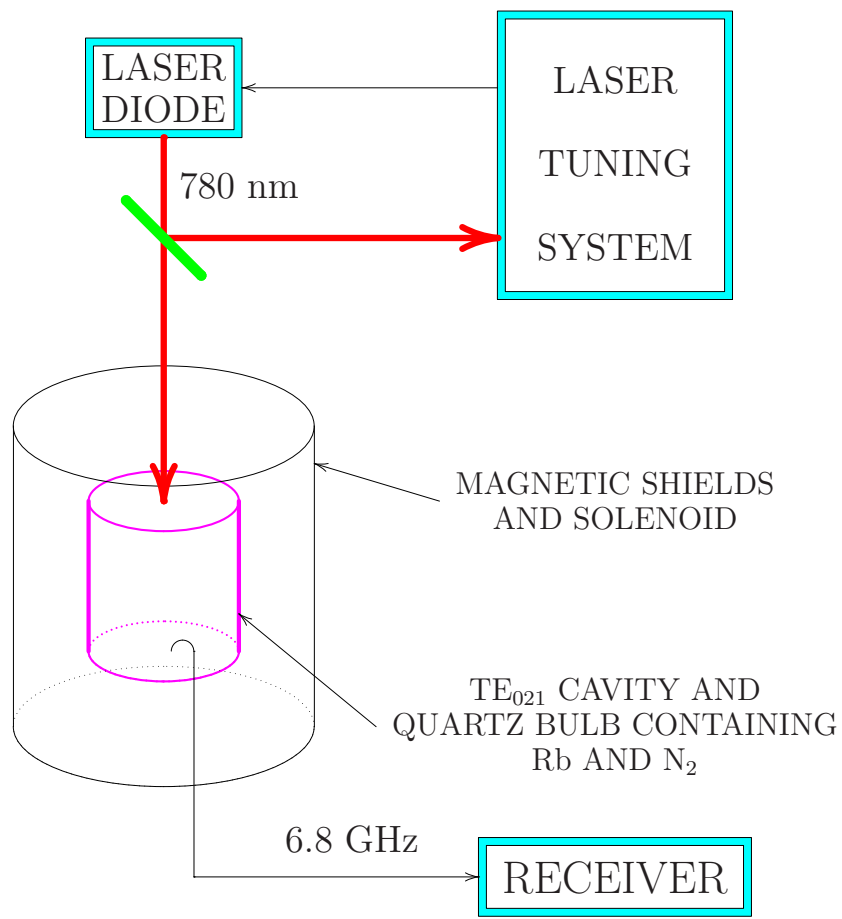

\[ \xy 
(0,65)*+[F**:white]+[F**:cyan]{\txt\large {LASER \\ DIODE}}="bld",
(60,56)*++++[F**:white]+[F**:cyan]{\txt\large {\\ LASER \\ \\ TUNING \\ \\ SYSTEM \\}}="blts",
(60,-40)*++[F**:white]+[F**:cyan]{\txt\Large {RECEIVER}}="breceiver";
 (0,5)*[magenta][|(2)]\ellipse(10,4){-}; 
 (0,-5)*[magenta][|(2)]\ellipse(10,4){.}; 
 (0,-5)*[magenta][|(2)]\ellipse(10,4)__,=:a(180){-}; 
 (-10,10)*{}="TL"; 
 (10,10)*{}="TR"; 
 (-10,-10)*{}="BL"; 
 (10,-10)*{}="BR"; 
 "TL"; "BL" **[magenta][|(4)]\dir{-}; 
 "TR"; "BR" **[magenta][|(4)]\dir{-};
 (0,10)*\ellipse(20,8){-}; 
 (0,-10)*\ellipse(20,8){.}; 
 (0,-10)*\ellipse(20,8),=:a(180){-}; 
 (-20,20)*{}="TL2"; 
 (20,20)*{}="TR2"; 
 (-20,-20)*{}="BL2"; 
 (20,-20)*{}="BR2"; 
 "TL2"; "BL2" **\dir{-}; 
 "TR2"; "BR2" **\dir{-};
(60,10)*++{\txt{MAGNETIC SHIELDS \\ AND SOLENOID}}="l1b",
(60,-20)*++{\txt{TE$_{021}$  CAVITY AND \\ QUARTZ BULB CONTAINING \\ Rb AND N$_{2}$}}="l2b",
"TR2"+(0,-5);"l1b"+L **@{-} 
?<*@{<};
"TR"+(0,-15);"l2b"+L **@{-} 
?<*@{<} ;
"bld"+R;"blts"+L+(0,9) **@{-} 
?<*@{<} ;
{\ar @{>} (4,-40)="corner" ;"breceiver"+L };
(4,-10)="loopend",
{\ar @{-}  "corner"; "loopend"};
"loopend"+(-3,0)="loopground",
"loopground",{\ellipse^{}};
{\ar@*{[red][|(6)]} "bld";(0,10)};
{\ar@*{[red][|(6)]}(0,45);"blts" + L + (0,-11) };
{\ar@*{[|(10)][green]}@{-} (-5,50) ; (5,40) };
(10,52)*{\txt\large {780 nm}};
(25,-35)*{\txt\large {6.8 GHz}}
\endxy \] 
\caption{Schematic of the laser diode pumped Rb maser.}
\label{fig:schema}
\end{figure}

The maser (Fig.~\ref{fig:schema}), uses a cylindrical
TE$_{021}$ mode microwave cavity
containing a quartz bulb in which a $^{87}$Rb vapor is present
with about 11 Torr of
N$_{2}$ used as a buffer gas. The cavity is made of a quartz cylinder coated
with silver and its resonance frequency is tuned to 6.83468~GHz
corresponding to
the transition frequency between the
two $m_{F} = 0$ ground state hyperfine
levels of the atoms.
One end of the cavity is equipped with a coupling loop to bring the maser
signal to the receiver, and with a piston for its fine frequency tuning.
The other end has a screen made of a silver coated corrugated ribbon
so the pumping light can enter into the cavity.
The loaded quality factor of
the cavity is around  27000.

The pumping light is derived from a commercial laser-diode selected for use at
780.027 nm, corresponding to the 5$^{2}$S$_{1/2}$, F=1 $\Longleftrightarrow$
5$^{2}$P$_{3/2},F=0,1,2$ transitions, the Doppler width of each transition
overlapping.
At this wavelength, the optical power
delivered by the laser is around 18 mW.

Coarse adjustment of the wavelength is
achieved by setting the case temperature of the laser with the use of a
thermoelectric cooler, while fine tuning is achieved with the change of the
injection current.  The light is first collimated and injected into the maser
cavity.
The collimated beam waist dimensions are about $3 \times 9$ mm.
The light enters the cavity near the side wall
and then shines in all directions and polarizations. It was not possible
with this setup to evaluate the light intensity distribution, but
this arrangement produced the best output power. Also, no optical isolator
was needed in order to avoid feedback light which could result in laser mode
hoppings. 

\begin{figure}
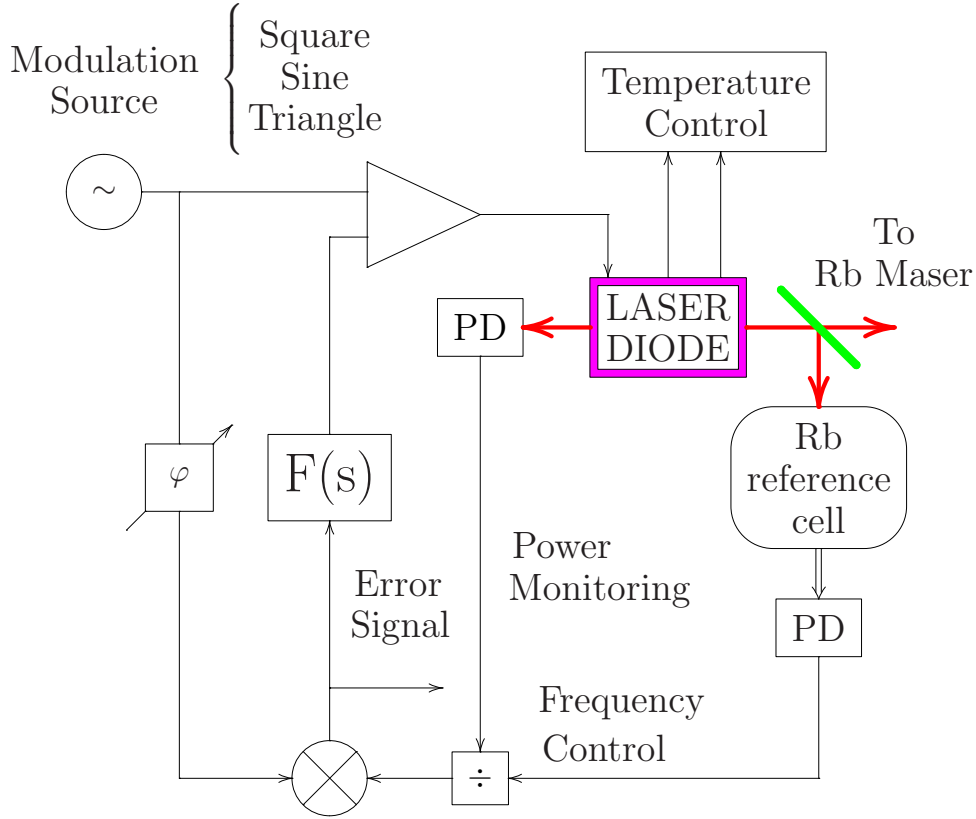

\[ \xy 
(10,30)*++[F]{\txt\large {PD}}="bpd";
(10,-30)*++[F]{\displaystyle \div}="db";
{\ar @{->}  "bpd"; "db"};
(35,30)*+[F**:white]+[F**:magenta]{\txt\large {LASER \\ DIODE}}="bld";
{\ar@*{[red][|(4)]} "bld";"bpd"};
(65,30)="outputbeam";
{\ar@*{[red][|(4)]} "bld";"outputbeam"};
(55,30)="mirrorposition";
"mirrorposition"+(0,-20)*++[F-:<16pt>]{\txt \large {Rb \\ reference \\ cell}}="rbcell"; 
{\ar@*{[red][|(4)]}"mirrorposition";"rbcell"};
{\ar@*{[|(8)][green]}@{-}"mirrorposition"+(-5,5);"mirrorposition"+(5,-5)};
"rbcell"+(0,-20)*++[F]{\txt\large {PD}}="pd2";  
{\ar @{=>} "rbcell";"pd2"};
(55,-30)="corner1";
{\ar @{-} "pd2";"corner1"};
{\ar @{->} "corner1";"db"+R};
{"db"+(-20,0)\ellipse(5){}}="mixer";
{\ar @{-} "mixer"+(3.5355,3.5355);"mixer"+(-3.5355,-3.5355)};
{\ar @{-} "mixer"+(3.5355,-3.5355);"mixer"+(-3.5355,3.5355)};
{\ar @{-} "mixer"+(0,5);"mixer"+(0,12)="errorsignal"};
{\ar @{->} "errorsignal";"errorsignal"+(15,0)}="outputerrorsignal";
"outputerrorsignal"+(5,26)*[r]{\txt \large {Error}};
"outputerrorsignal"+(5,20)*[r]{\txt \large {Signal}};
"outputerrorsignal"+(0,40)*++[F]{\txt\LARGE {F(s)}}="fb";
{\ar @{->} "errorsignal";"fb"};
"fb"+(-20,0)*+++[F]{\varphi}="phaseshifter";
{\ar @{->} "phaseshifter";"phaseshifter"+(7,7)};
{\ar @{-} "phaseshifter";"phaseshifter"+(-7,-7)};
{\ar @{<-} "mixer"+(-5,0);(-30,-30)="corner2"};
{\ar @{-} "corner2";"phaseshifter"+D};
{\ar @{-} "fb"+(20,35)="amplifieroutput";"amplifieroutput"+(-15,7)};
{\ar @{-} "amplifieroutput"; "amplifieroutput"+(-15,-7)};
{\ar @{-} "amplifieroutput"+(-15,7);"amplifieroutput"+(-15,-7)};
{"fb"+(-30,38)\ellipse(5){-}}="oscillator";
"oscillator"*{\displaystyle \sim};
{\ar @{-} "oscillator"+(5,0);"amplifieroutput"+(-15,3)};
{\ar @{-} "phaseshifter"+U;"oscillator"+(10,0)};
{\ar @{-} "fb"+U;"fb"+(0,32)="corner3"};
{\ar @{-} "corner3";"amplifieroutput"+(-15,-3)};
{\ar @{-} "amplifieroutput";"amplifieroutput"+(17,0)="corner4"};
{\ar @{->} "corner4";"bld"+U+(-8,0)};
"bld"+(5,30)*++[F]{\txt\large {Temperature\\ Control}}="btc";
{\ar @*{}"bld"+U;"btc"+D+(-5,0)};
{\ar @{->} "bld"+U+(7,0);"btc"+D+(2,0)};
"mirrorposition"+(10,10)*{\txt\large {To \\ Rb Maser}};
"pd2"+(-35,-10)*[r]{\txt\large {Frequency}};
"pd2"+(-35,-16)*[r]{\txt\large {Control}};
"db"+(6,31)*[r]{\txt\large {Power}};
"db"+(6,25)*[r]{\txt\large {Monitoring}};
{\ar @{->}  "db"+L; "mixer"+(5,0)};
"oscillator"+(0,15)*++{\txt\large{Modulation \\ Source}};
"oscillator"+(28,15)*++{\txt\large{Square \\ Sine \\ Triangle }}*\frm{\{};
\endxy \] 
\caption{Schematic of the laser locking system.}
\label{fig:electronique}
\end{figure}

A portion of the laser beam is sent through an external
cell containing also a Rb vapor
in order to lock the nominal laser frequency on the absorption resonance.
Fig.~\ref{fig:electronique} shows the arrangement
for the laser frequency-locking system.
An additional photodetector and an
analog voltage divider were used
to compensate for the change in the output power of the laser
when the injection current is varied.
  The signal from the divider is then phase compared with the modulation
signal.
 The resulting error signal is then proportional
to the frequency
difference between the laser and the absorption resonance.

\begin{figure}
%
\psset{xunit=4cm,yunit=0.6cm}
\begin{pspicture}(-1.8,-10)(1.5,8)
\psaxes[Ox=-1.5,Dx=0.5,Oy=-8,Dy=2,axesstyle=frame,tickstyle=top](-1.5,-8)(1.5,8)
\savedata{\mydata}[
{{-0.9348,-1.5633},
{-0.9232,-1.5659},
{-0.9115,-1.5342},
{-0.8998,-1.5688},
{-0.8881,-1.6407},
{-0.8764,-1.7498},
{-0.8647,-1.7652},
{-0.8530,-1.7785},
{-0.8414,-1.9629},
{-0.8297,-1.9704},
{-0.8180,-2.0715},
{-0.8063,-2.2246},
{-0.7946,-2.3097},
{-0.7829,-2.3461},
{-0.7712,-2.5503},
{-0.7596,-2.5742},
{-0.7479,-2.7760},
{-0.7362,-2.7570},
{-0.7245,-2.9121},
{-0.7128,-2.8990},
{-0.7011,-3.1003},
{-0.6895,-3.2106},
{-0.6778,-3.6641},
{-0.6661,-3.7035},
{-0.6544,-3.8555},
{-0.6427,-3.7897},
{-0.6310,-3.9195},
{-0.6193,-4.1343},
{-0.6077,-4.2505},
{-0.5960,-4.5071},
{-0.5843,-4.7275},
{-0.5726,-4.8765},
{-0.5609,-5.0730},
{-0.5492,-5.2290},
{-0.5375,-5.2162},
{-0.5259,-5.3508},
{-0.5142,-5.4425},
{-0.5025,-5.6461},
{-0.4908,-5.6756},
{-0.4791,-5.7511},
{-0.4674,-5.8386},
{-0.4558,-5.8689},
{-0.4441,-5.8851},
{-0.4324,-5.9268},
{-0.4207,-5.9260},
{-0.4090,-5.9344},
{-0.3973,-5.8815},
{-0.3856,-5.8150},
{-0.3740,-5.7819},
{-0.3623,-5.7050},
{-0.3506,-5.6339},
{-0.3389,-5.5629},
{-0.3272,-5.4417},
{-0.3155,-5.3800},
{-0.3038,-5.3670},
{-0.2922,-5.2615},
{-0.2805,-4.9979},
{-0.2688,-4.7874},
{-0.2571,-4.7603},
{-0.2454,-4.7118},
{-0.2337,-4.7384},
{-0.2220,-4.5611},
{-0.2104,-4.2933},
{-0.1987,-4.1362},
{-0.1870,-3.9946},
{-0.1753,-3.7996},
{-0.1636,-3.5335},
{-0.1519,-3.4208},
{-0.1403,-3.2512},
{-0.1286,-2.9798},
{-0.1169,-2.7636},
{-0.1052,-2.6468},
{-0.0935,-2.2465},
{-0.0818,-1.8821},
{-0.0701,-1.4790},
{-0.0585,-1.2453},
{-0.0468,-1.0833},
{-0.0351,-0.9550},
{-0.0234,-0.6252},
{-0.0117,-0.3351},
{0.0000,-0.0812},
{0.0117,0.3088},
{0.0233,0.2886},
{0.0350,0.6218},
{0.0467,0.8468},
{0.0584,1.0612},
{0.0701,1.1563},
{0.0818,1.2954},
{0.0934,1.6313},
{0.1051,1.6898},
{0.1168,1.7531},
{0.1285,1.9641},
{0.1402,2.0547},
{0.1519,2.4361},
{0.1636,2.7208},
{0.1752,2.9722},
{0.1869,2.9514},
{0.1986,3.2637},
{0.2103,3.6965},
{0.2220,3.8394},
{0.2337,3.9844},
{0.2454,4.0843},
{0.2570,4.3430},
{0.2687,4.4983},
{0.2804,4.5679},
{0.2921,4.7890},
{0.3038,4.8698},
{0.3155,4.9061},
{0.3272,5.0322},
{0.3388,5.3296},
{0.3505,5.5062},
{0.3622,5.6365},
{0.3739,5.7407},
{0.3856,5.8114},
{0.3973,5.9215},
{0.4089,5.9705},
{0.4206,6.0169},
{0.4323,6.0397},
{0.4440,6.0564},
{0.4557,6.0881},
{0.4674,6.1171},
{0.4791,6.1443},
{0.4907,6.1334},
{0.5024,6.1267},
{0.5141,6.1348},
{0.5258,6.0725},
{0.5375,6.0326},
{0.5492,5.9366},
{0.5609,5.8761},
{0.5725,5.7737},
{0.5842,5.7607},
{0.5959,5.7093},
{0.6076,5.5983},
{0.6193,5.3994},
{0.6310,5.3826},
{0.6426,5.2599},
{0.6543,5.1245},
{0.6660,5.0427},
{0.6777,4.8874},
{0.6894,4.8089},
{0.7011,4.6453},
{0.7128,4.4744},
{0.7244,4.3889},
{0.7361,4.1192},
{0.7478,3.9615},
{0.7595,3.8512},
{0.7712,3.6996},
{0.7829,3.3845},
{0.7946,3.1974},
{0.8062,2.9043},
{0.8179,2.8104},
{0.8296,2.6385},
{0.8413,2.4743},
{0.8530,2.5041},
{0.8647,2.1813},
{0.8764,2.0104},
{0.8880,1.8825},
{0.8997,1.6201},
{0.9114,1.3719},
{0.9231,1.3822},
{0.9348,1.2884},
{0.9465,1.1734},
{0.9581,1.0403},
{0.9698,0.9707},
{0.9815,0.9141},
{0.9932,0.8277},
{1.0049,0.7038},
{1.0166,0.6461},
{1.0283,0.5387},
{1.0399,0.4523},
{1.0516,0.4146},
{1.0633,0.3851},
{1.0750,0.3353},
{1.0867,0.2729},
{1.0984,0.2257},
{1.1101,0.1738},
{1.1217,0.1149},
{1.1334,0.0832},
{1.1451,0.0940},
{1.1568,0.0826},
{1.1685,0.0402},
{1.1802,0.0235},
{1.1918,0.0298},
{1.2035,0.0032},
{1.2152,-0.0145},
{1.2269,-0.0297},
{1.2386,-0.0397},
{1.2503,-0.0508},
{1.2620,-0.0587},
{1.2736,-0.0718},
{1.2853,-0.0741},
{1.2970,-0.0798},
{1.3087,-0.0898},
{1.3204,-0.0900},
{1.3321,-0.0966},
{1.3438,-0.0991},
{1.3554,-0.0990},
{1.3671,-0.1026},
{1.3788,-0.1067}}]
\dataplot[plotstyle=dots,showpoints=true,
   dotstyle=+,dotscale=1.0,linecolor=green]{\mydata}
\savedata{\mydata}[{
{-1.4840,-1.4631},
{-1.4723,-1.5750},
{-1.4606,-1.6408},
{-1.4489,-1.6773},
{-1.4373,-1.7633},
{-1.4256,-1.8512},
{-1.4139,-1.8687},
{-1.4022,-1.8315},
{-1.3905,-2.0023},
{-1.3788,-2.1242},
{-1.3672,-2.2495},
{-1.3555,-2.3912},
{-1.3438,-2.3265},
{-1.3321,-2.4135},
{-1.3204,-2.4099},
{-1.3087,-2.5433},
{-1.2970,-2.6969},
{-1.2854,-2.8109},
{-1.2737,-2.9583},
{-1.2620,-3.0640},
{-1.2503,-3.2578},
{-1.2386,-3.2566},
{-1.2269,-3.2373},
{-1.2152,-3.2157},
{-1.2036,-3.3004},
{-1.1919,-3.2377},
{-1.1802,-3.4002},
{-1.1685,-3.5105},
{-1.1568,-3.6492},
{-1.1451,-3.7442},
{-1.1335,-3.8992},
{-1.1218,-4.0819},
{-1.1101,-4.2225},
{-1.0984,-4.1894},
{-1.0867,-4.1800},
{-1.0750,-4.2602},
{-1.0633,-4.4715},
{-1.0517,-4.5827},
{-1.0400,-4.7819},
{-1.0283,-4.9103},
{-1.0166,-4.8497},
{-1.0049,-4.9943},
{-0.9932,-4.9481},
{-0.9815,-5.0115},
{-0.9699,-5.0747},
{-0.9582,-5.0736},
{-0.9465,-5.2511},
{-0.9348,-5.2080},
{-0.9231,-5.1554},
{-0.9114,-5.2086},
{-0.8997,-5.2924},
{-0.8881,-5.3274},
{-0.8764,-5.3630},
{-0.8647,-5.2948},
{-0.8530,-5.4781},
{-0.8413,-5.5520},
{-0.8296,-5.6215},
{-0.8180,-5.6738},
{-0.8063,-5.7233},
{-0.7946,-5.7080},
{-0.7829,-5.7146},
{-0.7712,-5.8029},
{-0.7595,-5.8202},
{-0.7478,-5.9029},
{-0.7362,-6.0121},
{-0.7245,-5.8936},
{-0.7128,-5.9392},
{-0.7011,-5.9324},
{-0.6894,-5.9761},
{-0.6777,-5.9748},
{-0.6660,-5.9478},
{-0.6544,-6.0121},
{-0.6427,-5.9508},
{-0.6310,-5.9162},
{-0.6193,-5.9149},
{-0.6076,-5.8546},
{-0.5959,-5.8107},
{-0.5843,-5.7846},
{-0.5726,-5.7389},
{-0.5609,-5.6783},
{-0.5492,-5.6094},
{-0.5375,-5.4694},
{-0.5258,-5.4839},
{-0.5141,-5.5539},
{-0.5025,-5.4744},
{-0.4908,-5.7822},
{-0.4791,-5.6620},
{-0.4674,-5.5212},
{-0.4557,-5.4553},
{-0.4440,-5.4178},
{-0.4323,-5.3150},
{-0.4207,-5.1725},
{-0.4090,-5.0771},
{-0.3973,-4.9640},
{-0.3856,-4.7333},
{-0.3739,-4.7643},
{-0.3622,-4.6123},
{-0.3505,-4.4868},
{-0.3389,-4.3608},
{-0.3272,-4.2096},
{-0.3155,-4.0473},
{-0.3038,-3.9674},
{-0.2921,-3.5872}
{-0.2804,-3.4236},
{-0.2688,-3.1331},
{-0.2571,-3.1906},
{-0.2454,-2.9215},
{-0.2337,-2.7025},
{-0.2220,-2.4089},
{-0.2103,-2.3201},
{-0.1986,-2.2082},
{-0.1870,-2.0432},
{-0.1753,-1.8583},
{-0.1636,-1.8773},
{-0.1519,-1.6297},
{-0.1402,-1.4511},
{-0.1285,-1.2845},
{-0.1168,-1.1206},
{-0.1052,-1.0544},
{-0.0935,-1.0375},
{-0.0818,-0.9887},
{-0.0701,-0.8905},
{-0.0584,-0.6296},
{-0.0467,-0.4202},
{-0.0351,-0.2408},
{-0.0234,-0.1569},
{-0.0117,-0.2055},
{0.0000,-0.0046},
{0.0117,0.0653},
{0.0234,0.2449},
{0.0351,0.4868},
{0.0467,0.6685},
{0.0584,0.7108},
{0.0701,0.8617},
{0.0818,1.1340},
{0.0935,1.2815},
{0.1052,1.3207},
{0.1169,1.4569},
{0.1285,1.6540},
{0.1402,1.8522},
{0.1519,2.1426},
{0.1636,2.2252},
{0.1753,2.3818},
{0.1870,2.5701},
{0.1987,2.6533},
{0.2103,2.7787},
{0.2220,3.0979},
{0.2337,3.3281},
{0.2454,3.4583},
{0.2571,3.7104},
{0.2688,3.8478},
{0.2804,4.0134},
{0.2921,4.1335},
{0.3038,4.2765},
{0.3155,4.5275},
{0.3272,4.6722},
{0.3389,4.6800},
{0.3506,4.7536},
{0.3622,4.8124},
{0.3739,5.0067},
{0.3856,5.2851},
{0.3973,5.4191},
{0.4090,5.5339},
{0.4207,5.6631},
{0.4324,5.7182},
{0.4440,5.8139},
{0.4557,5.8018},
{0.4674,5.9089},
{0.4791,5.9802},
{0.4908,6.0472},
{0.5025,6.0867},
{0.5141,6.1485},
{0.5258,6.1956},
{0.5375,6.2087},
{0.5492,6.2160},
{0.5609,6.2163},
{0.5726,6.2244},
{0.5843,6.1991},
{0.5959,6.1687},
{0.6076,6.1520},
{0.6193,6.0992},
{0.6310,6.0393},
{0.6427,5.9605},
{0.6544,5.8953},
{0.6661,5.8370},
{0.6777,5.7963},
{0.6894,5.6779},
{0.7011,5.5087}}]
\dataplot[plotstyle=dots,showpoints=true,dotstyle=+,dotscale=1.0,linecolor=blue]{\mydata}
\savedata{\mydata}[{
{-1.0400,0.6896},
{-1.0283,0.7473},
{-1.0166,0.7807},
{-1.0049,0.7592},
{-0.9932,0.7739},
{-0.9815,0.8053},
{-0.9698,0.8729},
{-0.9582,0.9227},
{-0.9465,1.0230},
{-0.9348,1.0226},
{-0.9231,1.0829},
{-0.9114,1.1217},
{-0.8997,1.1872},
{-0.8880,1.2008},
{-0.8764,1.2607},
{-0.8647,1.2837},
{-0.8530,1.3074},
{-0.8413,1.4224},
{-0.8296,1.4843},
{-0.8179,1.5737},
{-0.8063,1.6882},
{-0.7946,1.7732},
{-0.7829,1.8427},
{-0.7712,1.9542},
{-0.7595,1.9161},
{-0.7478,1.9916},
{-0.7361,2.0406},
{-0.7245,2.0878},
{-0.7128,2.1487},
{-0.7011,2.1363},
{-0.6894,2.1499},
{-0.6777,2.1735},
{-0.6660,2.1702},
{-0.6543,2.1820},
{-0.6427,2.1770},
{-0.6310,2.1545},
{-0.6193,2.1452},
{-0.6076,2.1428},
{-0.5959,2.1561},
{-0.5842,2.0642},
{-0.5726,1.9697},
{-0.5609,1.9016},
{-0.5492,1.9023},
{-0.5375,1.8799},
{-0.5258,1.8486},
{-0.5141,1.8330},
{-0.5024,1.8059},
{-0.4908,1.7454},
{-0.4791,1.6524},
{-0.4674,1.5304},
{-0.4557,1.5407},
{-0.4440,1.4036},
{-0.4323,1.3900},
{-0.4206,1.2271},
{-0.4090,1.1375},
{-0.3973,0.9773},
{-0.3856,0.9026},
{-0.3739,0.7095},
{-0.3622,0.6401},
{-0.3505,0.5328},
{-0.3388,0.3739},
{-0.3272,0.2046},
{-0.3155,0.1496},
{-0.3038,0.0411},
{-0.2921,-0.0001},
{-0.2804,-0.0371},
{-0.2687,-0.1476},
{-0.2571,-0.1319},
{-0.2454,-0.1794},
{-0.2337,-0.2539},
{-0.2220,-0.2769},
{-0.2103,-0.3110},
{-0.1986,-0.3503},
{-0.1869,-0.3495},
{-0.1753,-0.3955},
{-0.1636,-0.3961},
{-0.1519,-0.3894},
{-0.1402,-0.3922},
{-0.1285,-0.3663},
{-0.1168,-0.3502},
{-0.1051,-0.3179},
{-0.0935,-0.3153},
{-0.0818,-0.2833},
{-0.0701,-0.2801},
{-0.0584,-0.2243},
{-0.0467,-0.1562},
{-0.0350,-0.1247},
{-0.0234,-0.0506},
{-0.0117,-0.0268},
{0.0,0.0098},
{0.0117,0.0675},
{0.0234,0.1549},
{0.0351,0.2143},
{0.0468,0.3235},
{0.0584,0.4188},
{0.0701,0.4558},
{0.0818,0.4832},
{0.0935,0.4892},
{0.1052,0.5487},
{0.1169,0.5912},
{0.1286,0.6290},
{0.1402,0.6596},
{0.1519,0.6768},
{0.1636,0.6957},
{0.1753,0.7188},
{0.1870,0.7154},
{0.1987,0.7055},
{0.2104,0.6942},
{0.2220,0.6646},
{0.2337,0.6482},
{0.2454,0.6070},
{0.2571,0.4916},
{0.2688,0.4524},
{0.2805,0.3556},
{0.2921,0.2835},
{0.3038,0.1528},
{0.3155,0.0903},
{0.3272,0.0567},
{0.3389,-0.0233},
{0.3506,-0.1723},
{0.3623,-0.3220},
{0.3739,-0.4010},
{0.3856,-0.6728},
{0.3973,-0.7769},
{0.4090,-0.8615},
{0.4207,-0.9410},
{0.4324,-1.0712},
{0.4441,-1.1748},
{0.4557,-1.2830},
{0.4674,-1.3654},
{0.4791,-1.4860},
{0.4908,-1.6844},
{0.5025,-1.7740},
{0.5142,-1.8497},
{0.5258,-2.0271},
{0.5375,-2.0830},
{0.5492,-2.2230},
{0.5609,-2.2787},
{0.5726,-2.2960},
{0.5843,-2.4054},
{0.5960,-2.4564},
{0.6076,-2.4885},
{0.6193,-2.5109},
{0.6310,-2.4915},
{0.6427,-2.4802},
{0.6544,-2.4553},
{0.6661,-2.3951},
{0.6778,-2.3285},
{0.6894,-2.2440},
{0.7011,-2.1529},
{0.7128,-2.1364},
{0.7245,-2.0780},
{0.7362,-2.0585},
{0.7479,-1.9253},
{0.7596,-1.8556},
{0.7712,-1.7936},
{0.7829,-1.6420},
{0.7946,-1.4498},
{0.8063,-1.3419},
{0.8180,-1.2928},
{0.8297,-1.2295},
{0.8413,-1.1885},
{0.8530,-1.1110},
{0.8647,-0.9559},
{0.8764,-0.8596},
{0.8881,-0.7232},
{0.8998,-0.6948},
{0.9115,-0.6922},
{0.9231,-0.5863},
{0.9348,-0.5149},
{0.9465,-0.4633},
{0.9582,-0.4596},
{0.9699,-0.4137},
{0.9816,-0.3634},
{0.9933,-0.2635},
{1.0049,-0.2695},
{1.0166,-0.1946},
{1.0283,-0.1766},
{1.0400,-0.1535},
{1.0517,-0.1374},
{1.0634,-0.1306},
{1.0750,-0.1112},
{1.0867,-0.1178},
{1.0984,-0.0980},
{1.1101,-0.1149},
{1.1218,-0.0759},
{1.1335,-0.0607},
{1.1452,-0.0539},
{1.1568,-0.0410},
{1.1685,-0.0303},
{1.1802,-0.0221},
{1.1919,-0.0177},
{1.2036,-0.0057},
{1.2153,-0.0056},
{1.2270,0.0022},
{1.2386,0.0026},
{1.2503,0.0085},
{1.2620,0.0072}}]
\dataplot[plotstyle=dots,showpoints=true,dotstyle=+,dotscale=1.0,linecolor=red]{\mydata}
\rput*[l](0.8,-3){SQUARE}
\rput*[l](0.8,6){SINE}
\rput*[l](0.8,4){TRIANGLE}
\rput*[c](0.0,-10){LASER DETUNING [GHz]}
\rput*[c]{L}(-1.8,0.0){ERROR SIGNAL [V]}
\end{pspicture}
\caption{Error signal as a function of the average laser detuning for different modulation waveforms.}
\label{fig:pat}
\end{figure}
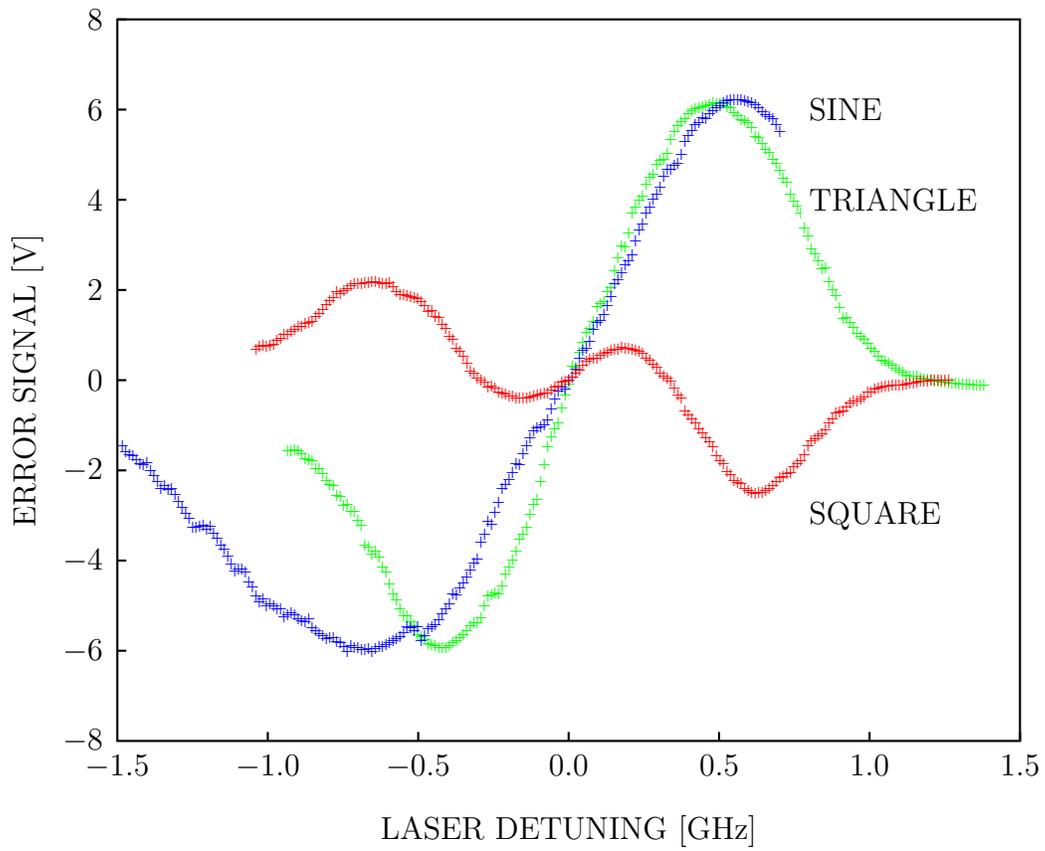

Three deterministic types of modulation signal were used: sine, triangle
and square waves.
Fig.~\ref{fig:pat}
shows the error signal observed when the laser is scanned across the
resonance line, for typical operating conditions
and for the different modulation waveforms.
The scale reference for the frequency axis is determined by measuring
the injection current required to reach the center of
each of the two absorption lines of the external cell which corresponds
to a 6.521 GHz frequency offset \cite{TET89}.
All the curves are adjusted for the coincidence of the
zero of the error signal. The same setting, frequency scale and
operating conditions are used
in Figs. \ref{fig:pat}, \ref{fig:puis}, \ref{fig:spec} and \ref{fig:freq}.

The discriminator gain and the locking range are about the same for both sine
and triangle wave modulation. They are somewhat restrained
for square wave,
due to the large modulation excursion. This should be taken into
account if this type of modulation is to be used.

\section{Output Power}

When the laser is unmodulated, its linewidth of about 25 MHz is too
narrow to pump a sufficiently large number of atoms and to maintain
the maser oscillation. Therefore it is necessary to modulate
the laser frequency. This modulation results in a broadening of the
laser spectrum, thus allowing optical pumping of all classes
of atomic velocities.
The benefits
of this broadening are limited by available laser power and undesired
pumping from
the 5S$_{1/2}$, F=2 level, thus resulting in a decrease
in the maser power.

Fig.~\ref{fig:puis}
shows the output power of the maser as a function of the average
laser frequency detuning
for the various modulation waveforms.
The amplitude of the modulation was adjusted for the
maximum maser power for each modulation waveform.
We also used a 50 kHz bandwidth,
Gaussian filtered white noise as a modulation source.
The {ero of the horizontal axis is arbitrary
for this type of modulation since the
locking system is useless in the absence of a periodic waveform.
All the curves are shown for the same optical power (18 mW) and modulation
frequency
(5 kHz).
We observe that the power is maximum when the modulating waveform is square.
On the opposite,
the Gaussian noise is the less efficient source of modulation.
We also note an assymetry of the curves. This is due to the presence of
the
neighboring 5S$_{1/2}$, F=2 $\Longleftrightarrow$
5P$_{3/2}$, F=1,2,3 transitions.

\begin{figure}
%
\psset{yunit=4cm,xunit=12cm}
\begin{pspicture}(-0.6,-0.5)(0.6,3.5)
\psaxes[Ox=-0.5,Dx=0.1,Dy=0.5,axesstyle=frame,tickstyle=top](-0.5,0)(0.5,3)
\savedata{\mydata}[
{{-2.506E-1,2.19087424E-2},
{-2.389E-1,2.860623360E-2},
{-2.272E-1,5.872224064E-1},
{-2.155E-1,6.473863104E-1},
{-2.038E-1,9.142643072E-1},
{-1.921E-1,9.851555488E-1},
{-1.804E-1,1.055025139E+0},
{-1.688E-1,1.203221322E+0},
{-1.571E-1,1.305159408E+0},
{-1.454E-1,1.350736403E+0},
{-1.337E-1,1.408402938E+0},
{-1.220E-1,1.494959498E+0},
{-1.103E-1,1.531455149E+0},
{-9.860E-2,1.575499667E+0},
{-8.700E-2,1.571242787E+0},
{-7.530E-2,1.609952016E+0},
{-6.360E-2,1.634925712E+0},
{-5.190E-2,1.648093661E+0},
{-4.020E-2,1.664042771E+0},
{-2.850E-2,1.672329498E+0},
{-1.690E-2,1.678402646E+0},
{-5.200E-3,1.684419037E+0},
{6.5000E-3,1.690832736E+0},
{1.8200E-2,1.685383930E+0},
{2.9900E-2,1.670286195E+0},
{4.1600E-2,1.658593965E+0},
{5.3300E-2,1.647526077E+0},
{6.4900E-2,1.631066141E+0},
{7.6600E-2,1.609611466E+0},
{8.8300E-2,1.594684006E+0},
{1.0000E-1,1.543147379E+0},
{1.1170E-1,1.561196550E+0},
{1.2340E-1,1.478726595E+0},
{1.3510E-1,1.438598406E+0},
{1.4670E-1,1.431163056E+0},
{1.5840E-1,1.334049434E+0},
{1.7010E-1,1.216389270E+0},
{1.8180E-1,1.184831600E+0},
{1.9350E-1,1.102475162E+0},
{2.0520E-1,9.749390368E-1},
{2.1690E-1,8.892906112E-1},
{2.2850E-1,7.162910080E-1},
{2.4020E-1,4.785868288E-1},
{2.5190E-1,4.623539264E-1}
{2.6360E-1,2.066573344E-1},
{2.7530E-1,2.066005760E-2}}]
\dataplot[plotstyle=dots,showpoints=true,
      dotstyle=+,dotscale=1.6,
      linecolor=black]{\mydata}
\savedata{\mydata}[
{{-3.038E-1,2.100060800E-2},
{-2.922E-1,3.825516160E-2},
{-2.805E-1,7.804280000E-1},
{-2.688E-1,1.003545270E+0},
{-2.571E-1,1.075344646E+0},
{-2.454E-1,1.153160413E+0},
{-2.337E-1,1.136586960E+0},
{-2.220E-1,1.271615194E+0},
{-2.104E-1,1.422422262E+0},
{-1.987E-1,1.491383718E+0},
{-1.870E-1,1.544623098E+0},
{-1.753E-1,1.614946755E+0},
{-1.636E-1,1.703376342E+0},
{-1.519E-1,1.726477011E+0},
{-1.403E-1,1.784370579E+0},
{-1.286E-1,1.841412771E+0},
{-1.169E-1,1.873651542E+0},
{-1.052E-1,1.906060589E+0},
{-9.350E-2,1.942726515E+0},
{-8.180E-2,1.995965894E+0},
{-7.010E-2,2.033539955E+0},
{-5.850E-2,2.044778118E+0},
{-4.680E-2,2.049091757E+0},
{-3.510E-2,2.051418851E+0},
{-2.340E-2,2.062827290E+0},
{-1.170E-2,2.072703251E+0},
{0.000E+0,2.075030346E+0},
{1.170E-2,2.068219338E+0},
{2.330E-2,2.064530042E+0},
{3.500E-2,2.053178362E+0},
{4.670E-2,2.042167232E+0},
{5.840E-2,2.027183014E+0},
{7.010E-2,2.017250294E+0},
{8.180E-2,2.006466198E+0},
{9.340E-2,1.965713667E+0},
{1.051E-1,1.960151344E+0},
{1.168E-1,1.954191712E+0},
{1.285E-1,1.926380096E+0},
{1.402E-1,1.907195757E+0},
{1.519E-1,1.845272342E+0},
{1.636E-1,1.777048746E+0},
{1.752E-1,1.707689981E+0},
{1.869E-1,1.687938058E+0},
{1.986E-1,1.585772938E+0},
{2.103E-1,1.394440371E+0},
{2.220E-1,1.303626931E+0},
{2.337E-1,1.209918813E+0},
{2.454E-1,1.119956749E+0},
{2.570E-1,8.262887872E-1},
{2.687E-1,5.018577728E-1},
{2.804E-1,3.072332192E-1},
{2.921E-1,2.400880320E-2},
{3.038E-1,2.077357440E-2}}]
\dataplot[plotstyle=dots,showpoints=true,
   dotstyle=+,dotscale=1.6,linecolor=green]{\mydata}
\savedata{\mydata}[
{{-3.622E-1,2.088709120E-2},
{-3.505E-1,3.036574400E-2},
{-3.389E-1,1.200440160E-1},
{-3.272E-1,7.685087360E-1},
{-3.155E-1,8.968962368E-1},
{-3.038E-1,9.701145728E-1},
{-2.921E-1,1.182277472E+0},
{-2.804E-1,1.311005523E+0},
{-2.688E-1,1.437406480E+0},
{-2.571E-1,1.416632906E+0},
{-2.454E-1,1.550298938E+0},
{-2.337E-1,1.969232688E+0},
{-2.220E-1,1.755083245E+0},
{-2.103E-1,1.781986726E+0},
{-1.986E-1,1.819787821E+0},
{-1.870E-1,1.869394662E+0},
{-1.753E-1,1.922520525E+0},
{-1.636E-1,1.923144867E+0},
{-1.519E-1,1.992276598E+0},
{-1.402E-1,2.032972371E+0},
{-1.285E-1,2.069184230E+0},
{-1.168E-1,2.105679882E+0},
{-1.052E-1,2.122139818E+0},
{-9.350E-2,2.127702141E+0},
{-8.180E-2,2.139905197E+0},
{-7.010E-2,2.156762442E+0},
{-5.840E-2,2.197174422E+0},
{-4.670E-2,2.227540166E+0},
{-3.510E-2,2.250413802E+0},
{-2.340E-2,2.262616858E+0},
{-1.170E-2,2.257622118E+0},
{0.000E+0,2.276352390E+0},
{1.170E-2,2.280893062E+0},
{2.340E-2,2.292869085E+0},
{3.510E-2,2.306604618E+0},
{4.670E-2,2.311088531E+0},
{5.840E-2,2.313983210E+0},
{7.010E-2,2.313415626E+0},
{8.180E-2,2.305015382E+0},
{9.350E-2,2.292698810E+0},
{1.052E-1,2.293436669E+0},
{1.169E-1,2.280495754E+0},
{1.285E-1,2.260857347E+0},
{1.402E-1,2.233329523E+0},
{1.519E-1,2.182360480E+0},
{1.636E-1,2.164424826E+0},
{1.753E-1,2.126453456E+0},
{1.870E-1,2.073157318E+0},
{1.987E-1,2.054086496E+0},
{2.103E-1,2.014412374E+0},
{2.220E-1,1.885116739E+0},
{2.337E-1,1.767513334E+0},
{2.454E-1,1.698268086E+0},
{2.571E-1,1.532874109E+0},
{2.688E-1,1.425998042E+0},
{2.804E-1,1.262987917E+0},
{2.921E-1,1.131251670E+0},
{3.038E-1,9.495112736E-1},
{3.155E-1,4.074685536E-1},
{3.272E-1,5.006090880E-2},
{3.389E-1,2.196550080E-2}}]
\dataplot[plotstyle=dots,showpoints=true,
  dotstyle=+,dotscale=1.6,
  linecolor=blue]{\mydata}
\savedata{\mydata}[
{{-3.1550E-1,2.338446080E-2},
{-3.038E-1,6.757655104E-1},
{-2.921E-1,9.184644288E-1},
{-2.804E-1,1.026078355E+0},
{-2.687E-1,1.294545587E+0},
{-2.571E-1,1.294886138E+0},
{-2.454E-1,1.385359027E+0},
{-2.337E-1,1.595592141E+0},
{-2.220E-1,1.672556531E+0},
{-2.103E-1,1.786640915E+0},
{-1.986E-1,1.890054720E+0},
{-1.869E-1,1.930693734E+0},
{-1.753E-1,2.092171382E+0},
{-1.636E-1,2.155229965E+0},
{-1.519E-1,2.270449517E+0},
{-1.402E-1,2.304050490E+0},
{-1.285E-1,2.423243130E+0},
{-1.168E-1,2.456844102E+0},
{-1.051E-1,2.502875165E+0},
{-9.350E-2,2.517462074E+0},
{-8.180E-2,2.557136195E+0},
{-7.010E-2,2.554922618E+0},
{-5.840E-2,2.601123955E+0},
{-4.670E-2,2.645338749E+0},
{-3.500E-2,2.665374464E+0},
{-2.340E-2,2.699088954E+0},
{-1.170E-2,2.706921613E+0},
{0.000E+0,2.720940938E+0},
{1.170E-2,2.730476349E+0},
{2.340E-2,2.748241728E+0},
{3.510E-2,2.756982522E+0},
{4.680E-2,2.762828637E+0},
{5.840E-2,2.756017629E+0},
{7.010E-2,2.751420198E+0},
{8.180E-2,2.743757814E+0},
{9.350E-2,2.742395613E+0},
{1.052E-1,2.721281488E+0},
{1.169E-1,2.710894701E+0},
{1.286E-1,2.692448221E+0},
{1.402E-1,2.661685168E+0},
{1.519E-1,2.623089456E+0},
{1.636E-1,2.594653498E+0},
{1.753E-1,2.532900358E+0},
{1.870E-1,2.400823562E+0},
{1.987E-1,2.335948710E+0},
{2.104E-1,2.312110182E+0},
{2.220E-1,2.209377478E+0},
{2.337E-1,2.154946173E+0},
{2.454E-1,2.027183014E+0},
{2.571E-1,1.729541965E+0},
{2.688E-1,1.619260394E+0},
{2.805E-1,1.348466067E+0},
{2.921E-1,1.061609114E+0},
{3.038E-1,4.243257984E-1},
{3.155E-1,1.123248736E-1},
{3.272E-1,1.929785600E-2}}]
\dataplot[plotstyle=dots,showpoints=true,
  dotstyle=+,dotscale=1.6,
  linecolor=red]{\mydata}
\put(3.5,11){SQUARE}
\rput*[l]{-145}(0.27,2.78){\psline[linewidth=0.2pt]{-}(0.12,0)}
\put(3.5,10){SINE}
\rput*[l]{-145}(0.27,2.53){\psline[linewidth=0.1pt]{-}(0.16,0)}
\put(3.5,9){TRIANGLE}
\rput*[l]{-145}(0.27,2.29){\psline[linewidth=0.1pt]{-}(0.19,0)}
\put(3.5,8){NOISE}
\rput*[l]{-145}(0.27,2.04){\psline[linewidth=0.1pt]{-}(0.24,0)}
\put(-2.5,-1.2){LASER DETUNING [GHz]}
\rput*[c]{L}(-0.6,1.5){MASER OUTPUT POWER [pW]}
\end{pspicture}
\caption{Output power of the maser as a function of the average laser detuning for different modulation waveforms.}
\label{fig:puis}
\end{figure}
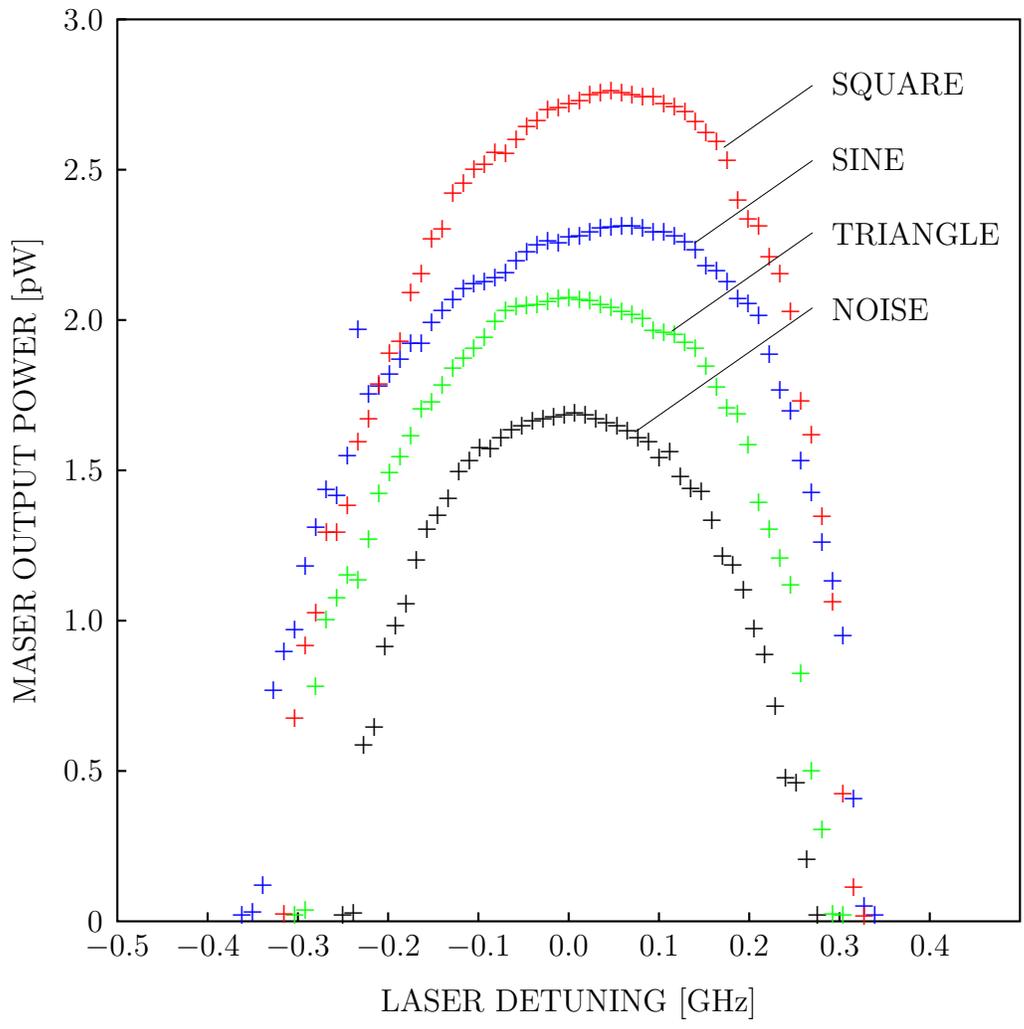

The laser spectrum for low modulation frequency can be expressed as the
convolution of the unmodulated laser spectrum
itself and the probability density function of the modulating signal.
For example, since a randomly starting triangle wave has a uniform
probability density function, the central part of the modulated
laser spectrum wiff also be uniform.
Fig.~\ref{fig:spec} shows the estimated spectrum of the laser
for the four modulation waveforms used in the experiments.
The modulation amplitudes are respectively 1.08, 1.40, 1.75 GHz peak
for square, sine and triangle and 1.3 GHz RMS for the gaussian noise.
It also shows the unmodulated laser spectrum and absorption curve
of the vapor, which is a Voigt profile.
\begin{figure}
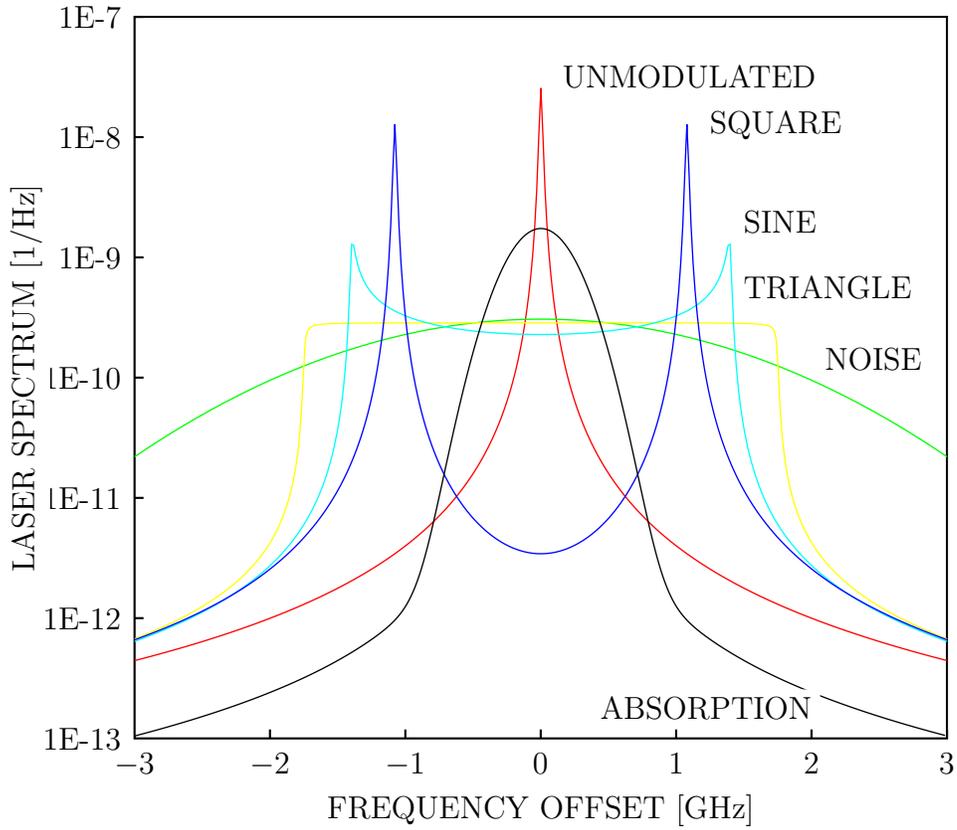

%
\psset{yunit=1.6cm,xunit=1.8cm}

\caption{Estimated light spectrum of the modulated laser for three deterministic modulation waveform, random noise modulation and the absorption spectrum of the atoms (black line).}
\label{fig:spec}
\end{figure}

This figure, along with Fig.~\ref{fig:puis} shows that the best maser output
power is obtained by
an optimum distribution of the light intensity
over the absorbtion profile of the atomic vapour.
Hence a narrow linewidth laser would cause excessive relaxation
for the microwave coherence, thus killing the maser action.

The maser output power also varies drastically with the modulation
frequency. Fig.~\ref{fig:pvsf} shows the output power as a function
of the modulation frequency. We see that the power suddenly drops
when the frequency is below a certain threshold. In fact, when the
modulation period comes close to the relaxation times, the changes in
laser frequency are not fast enough to pump all class of atoms
simultaneously and
the gain drops. This means that the
pumping of the atoms is not only due to the power spectrum of the
pumping light, but also to the speed of its variation.

\begin{figure}
\psset{yunit=0.4cm,xunit=4cm}
\begin{pspicture}(-0.3,-114)(3.5,-90)
\psaxes[Dx=1.0,Oy=-112,Dy=2,axesstyle=frame,
      tickstyle=top,labels=y,ticks=xy](0,-112)(0,-112)(3,-92)
\rput(0,-113){0.1}
\rput(1,-113){1}
\rput(2,-113){10}
\rput(3,-113){100}
\rput(1.5,-114.5){MODULATION FREQUENCY [kHz]}
\rput*[c]{L}(-0.4,-102){MASER OUTPUT POWER [dBm]}
\rput*[l](0.2,-94){SQUARE}
\rput*[l](0.2,-104){SINE}
\rput*[l](1,-106){TRIANGLE}
\psline[linewidth=0.1pt,linecolor=green](0.30103,-112)(0.30103,-92)
\psline[linewidth=0.1pt,linecolor=green](0.477121254,-112)(0.477121254,-92)
\psline[linewidth=0.1pt,linecolor=green](0.60206,-112)(0.60206,-92)
\psline[linewidth=0.1pt,linecolor=green](0.69897,-112)(0.69897,-92)
\psline[linewidth=0.1pt,linecolor=green](0.77815,-112)(0.77815,-92)
\psline[linewidth=0.1pt,linecolor=green](0.8451,-112)(0.8451,-92)
\psline[linewidth=0.1pt,linecolor=green](0.90309,-112)(0.90309,-92)
\psline[linewidth=0.1pt,linecolor=green](0.954242,-112)(0.954242,-92)
\psline[linewidth=0.1pt,linecolor=green](1,-112)(1,-92)
\psline[linewidth=0.1pt,linecolor=green](1.30103,-112)(1.30103,-92)
\psline[linewidth=0.1pt,linecolor=green](1.477121254,-112)(1.477121254,-92)
\psline[linewidth=0.1pt,linecolor=green](1.60206,-112)(1.60206,-92)
\psline[linewidth=0.1pt,linecolor=green](1.69897,-112)(1.69897,-92)
\psline[linewidth=0.1pt,linecolor=green](1.77815,-112)(1.77815,-92)
\psline[linewidth=0.1pt,linecolor=green](1.8451,-112)(1.8451,-92)
\psline[linewidth=0.1pt,linecolor=green](1.90309,-112)(1.90309,-92)
\psline[linewidth=0.1pt,linecolor=green](1.954242,-112)(1.954242,-92)
\psline[linewidth=0.1pt,linecolor=green](2,-112)(2,-92)
\psline[linewidth=0.1pt,linecolor=green](2.30103,-112)(2.30103,-92)
\psline[linewidth=0.1pt,linecolor=green](2.477121254,-112)(2.477121254,-92)
\psline[linewidth=0.1pt,linecolor=green](2.60206,-112)(2.60206,-92)
\psline[linewidth=0.1pt,linecolor=green](2.69897,-112)(2.69897,-92)
\psline[linewidth=0.1pt,linecolor=green](2.77815,-112)(2.77815,-92)
\psline[linewidth=0.1pt,linecolor=green](2.8451,-112)(2.8451,-92)
\psline[linewidth=0.1pt,linecolor=green](2.90309,-112)(2.90309,-92)
\psline[linewidth=0.1pt,linecolor=green](2.954242,-112)(2.954242,-92)
\savedata{\mydata}[{
{0.30103,-98.67},
{0.47712,-96.60},
{0.69897,-95.70},
{0.8451,-94.71},
{1.0,-94.75},
{1.30103,-94.05},
{1.47712,-93.82},
{1.69897,-93.75},
{1.8451,-93.85},
{2,-93.52},
{2.30103,-93.75}}]
\dataplot[plotstyle=curve,linewidth=2pt,linecolor=red]{\mydata}
\savedata{\mydata}[{
{0.30103,-98.67},
{0.47712,-96.60},
{0.69897,-95.70},
{0.8451,-94.71},
{1.0,-94.75},
{1.30103,-94.05},
{1.47712,-93.82},
{1.69897,-93.75},
{1.8451,-93.85},
{2,-93.52},
{2.30103,-93.75}}]
\dataplot[plotstyle=dots,showpoints=true,dotstyle=+,dotscale=2,linecolor=black]{\mydata}
\savedata{\mydata}[{
{0.47712,-107.6},
{0.69897,-99.75},
{0.84509,-99.40},
{1.0,    -98.95},
{1.30103,-98.47},
{1.47712,-98.30},
{1.69897,-98.20},
{1.8451, -97.85},
{2.0,    -97.95}}]
\dataplot[plotstyle=curve,linewidth=2pt,linecolor=blue]{\mydata}
\savedata{\mydata}[{
{0.47712,-107.7},
{0.69897,-99.75},
{0.84509,-99.40},
{1.0,    -98.95},
{1.30103,-98.47},
{1.47712,-98.30},
{1.69897,-98.20},
{1.8451, -97.85},
{2.0,    -97.95}}]
\dataplot[plotstyle=dots,showpoints=true,dotstyle=+,dotscale=2,linecolor=black]{\mydata}
\savedata{\mydata}[{
{0.69897,-110.75},
{0.8451,-105.3},
{1.0,-103.05},
{1.30103,-101.4},
{1.47712,-101.15},
{1.69897,-100.1},
{1.8451,-99.9},
{2.0,-99.81},
{2.30103,-99.3}}]
\dataplot[plotstyle=curve,linewidth=2pt,linecolor=green]{\mydata}
\savedata{\mydata}[{
{0.69897,-110.75},
{0.8451,-105.3},
{1.0,-103.05},
{1.30103,-101.4},
{1.47712,-101.15},
{1.69897,-100.1},
{1.8451,-99.9},
{2.0,-99.81},
{2.30103,-99.3}}]
\dataplot[plotstyle=dots,showpoints=true,dotstyle=+,dotscale=2,linecolor=black]{\mydata}
\end{pspicture}
\caption{Maser output power as a function of the laser modulation frequency.}
\label{fig:pvsf}
\end{figure}
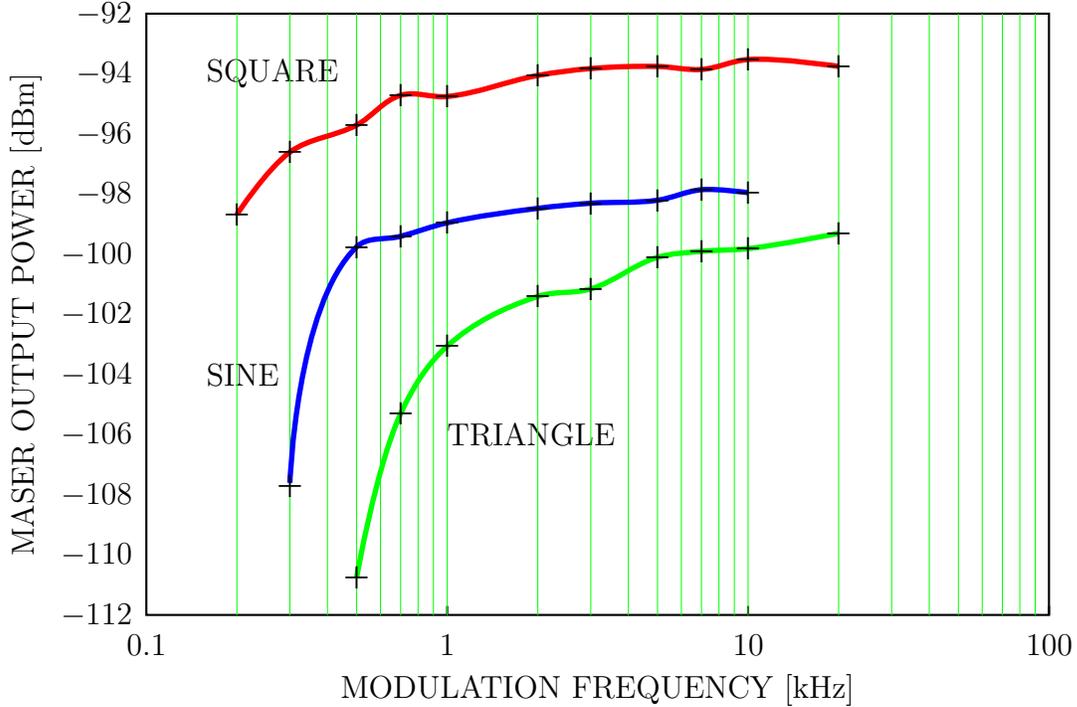

Fig.~\ref{fig:pvsi} shows the power of the maser when its temperature
and the light intensity are changed for square wave modulation.
We observe the same type of dependence as the
conventional maser. These curves show a maximum output power for
a certain light intensity. This maximum increases as the number of atoms
increases. Although this phenomenon will eventually be limited by relaxation
of the atoms, the behavior of the curves shows us that the output power of
the maser is actually limited by the available laser power. Other waveforms
showed a similar behavior although the maser was less powerful.

\begin{figure}
\psset{xunit=1cm,yunit=3cm}
\begin{pspicture}(5,-0.4)(18,3.6)
\psaxes[Ox=6,Dx=2.0,dx=0,Oy=0,Dy=0.5,axesstyle=frame,
      tickstyle=top,labels=all,ticks=all](6,0)(6,0)(18,3.5)
\rput(12,-0.4){LIGHT INTENSITY [mW]}
\rput*[c]{L}(4.75,1.75){MASER OUTPUT POWER [pW]}
\rput*[l](6.8,3.1){TEMPERATURE ($^{\circ}$C)}
\savedata{\mydata}[{
{7.1523,0.101157},
{8.08828,0.309029},
{9.05958,0.562341},
{10.61366,0.638263},
{12.50328,0.74131},
{14.51652,0.5652341},
{17.66,0.05248}}]
\dataplot[plotstyle=line,linewidth=2pt,linecolor=black]{\mydata}
\savedata{\mydata}[{
{7.1523,0.101157},
{8.08828,0.309029},
{9.05958,0.562341},
{10.61366,0.638263},
{12.50328,0.74131},
{14.51652,0.5652341},
{17.66,0.05248}}]
\dataplot[plotstyle=dots,showpoints=true,dotstyle=+,dotscale=2,linecolor=black]{\mydata}
\savedata{\mydata}[{{7.6,2.8},{8.4,2.8}}]
\dataplot[plotstyle=line,linewidth=2pt,linecolor=black]{\mydata}
\savedata{\mydata}[{8.0,2.8}]
\dataplot[plotstyle=dots,showpoints=true,dotstyle=+,dotscale=2,linecolor=black]{\mydata}
\rput*[l](9,2.8){54}
%
\savedata{\mydata}[{
{9.05958,0.147910},
{10.61366,0.451855},
{12.50328,0.851138},
{14.51652,1.049542},
{17.66,1.216186}}]
\dataplot[plotstyle=line,linewidth=2pt,linecolor=red]{\mydata}
\savedata{\mydata}[{
{9.05958,0.147910},
{10.61366,0.451855},
{12.50328,0.851138},
{14.51652,1.049542},
{17.66,1.216186}}]
\dataplot[plotstyle=dots,showpoints=true,dotstyle=+,dotscale=2,linecolor=red]{\mydata}
\savedata{\mydata}[{{7.6,2.6},{8.4,2.6}}]
\dataplot[plotstyle=line,linewidth=2pt,linecolor=red]{\mydata}
\savedata{\mydata}[{8.0,2.6}]
\dataplot[plotstyle=dots,showpoints=true,dotstyle=+,dotscale=2,linecolor=red]{\mydata}
\rput*[l](9,2.6){56}
%
\savedata{\mydata}[{
{9.059580,0.251188},
{10.61366,0.676082},
{12.50328,1.333521},
{14.51652,2.223309},
{17.66000,2.8119}}]
\dataplot[plotstyle=line,linewidth=2pt,linecolor=blue]{\mydata}
\savedata{\mydata}[{
{9.059580,0.251188},
{10.61366,0.676082},
{12.50328,1.333521},
{14.51652,2.223309},
{17.66000,2.8119}}]
\dataplot[plotstyle=dots,showpoints=true,dotstyle=+,dotscale=2,linecolor=blue]{\mydata}
\savedata{\mydata}[{
{7.6,2.4},
{8.4,2.4}}]
\dataplot[plotstyle=line,linewidth=2pt,linecolor=blue]{\mydata}
\savedata{\mydata}[{8.0,2.4}]
\dataplot[plotstyle=dots,showpoints=true,dotstyle=+,dotscale=2,linecolor=blue]{\mydata}
\rput*[l](9.0,2.4){58}
%
\savedata{\mydata}[{
{10.61366,0.038459},
{12.50328,1.135010},
{14.51652,2.018366},
{17.66000,2.884031}}]
\dataplot[plotstyle=line,linewidth=2pt,linecolor=green]{\mydata}
\savedata{\mydata}[{
{10.61366,0.038459},
{12.50328,1.135010},
{14.51652,2.018366},
{17.66000,2.884031}}]
\dataplot[plotstyle=dots,showpoints=true,dotstyle=+,dotscale=2,linecolor=green]{\mydata}
\savedata{\mydata}[{{7.6,2.2},{8.4,2.2}}]
\dataplot[plotstyle=line,linewidth=2pt,linecolor=green]{\mydata}
\savedata{\mydata}[{8.0,2.2}}]
\dataplot[plotstyle=dots,showpoints=true,dotstyle=+,dotscale=2,linecolor=green]{\mydata}
\rput*[l](9,2.2){60}
%
\savedata{\mydata}[{
{12.50328,0.402717},
{14.51652,1.496235},
{17.66,3.054921}}]
\dataplot[plotstyle=line,linewidth=2pt,linecolor=magenta]{\mydata}
\savedata{\mydata}[{
{12.50328,0.402717},
{14.51652,1.496235},
{17.66,3.054921}}]
\dataplot[plotstyle=dots,showpoints=true,dotstyle=+,dotscale=2,linecolor=magenta]{\mydata}
\savedata{\mydata}[{{7.6,2},{8.4,2}}]
\dataplot[plotstyle=line,linewidth=2pt,linecolor=magenta]{\mydata}
\savedata{\mydata}[{8.0,2}}]
\dataplot[plotstyle=dots,showpoints=true,dotstyle=+,dotscale=2,linecolor=magenta]{\mydata}
\rput*[l](9.0,2){61}
%
\savedata{\mydata}[{
{14.51652,0.245471},
{17.66,1.766037}}]
\dataplot[plotstyle=line,linewidth=2pt,linecolor=cyan]{\mydata}
\savedata{\mydata}[{
{14.51652,0.245471},
{17.66,1.766037}}]
\dataplot[plotstyle=dots,showpoints=true,dotstyle=+,dotscale=2,linecolor=cyan]{\mydata}
\savedata{\mydata}[{{7.6,1.8},{8.4,1.8}}]
\dataplot[plotstyle=line,linewidth=2pt,linecolor=cyan]{\mydata}
\savedata{\mydata}[{8.0,1.8}]
\dataplot[plotstyle=dots,showpoints=true,dotstyle=+,dotscale=2,linecolor=cyan]{\mydata}
\rput*[l](9.0,1.8){62}
\end{pspicture}
\caption{Maser power as a function of the light intensity for different temperatures and a square wave modulation of the laser.}
\label{fig:pvsi}
\end{figure}
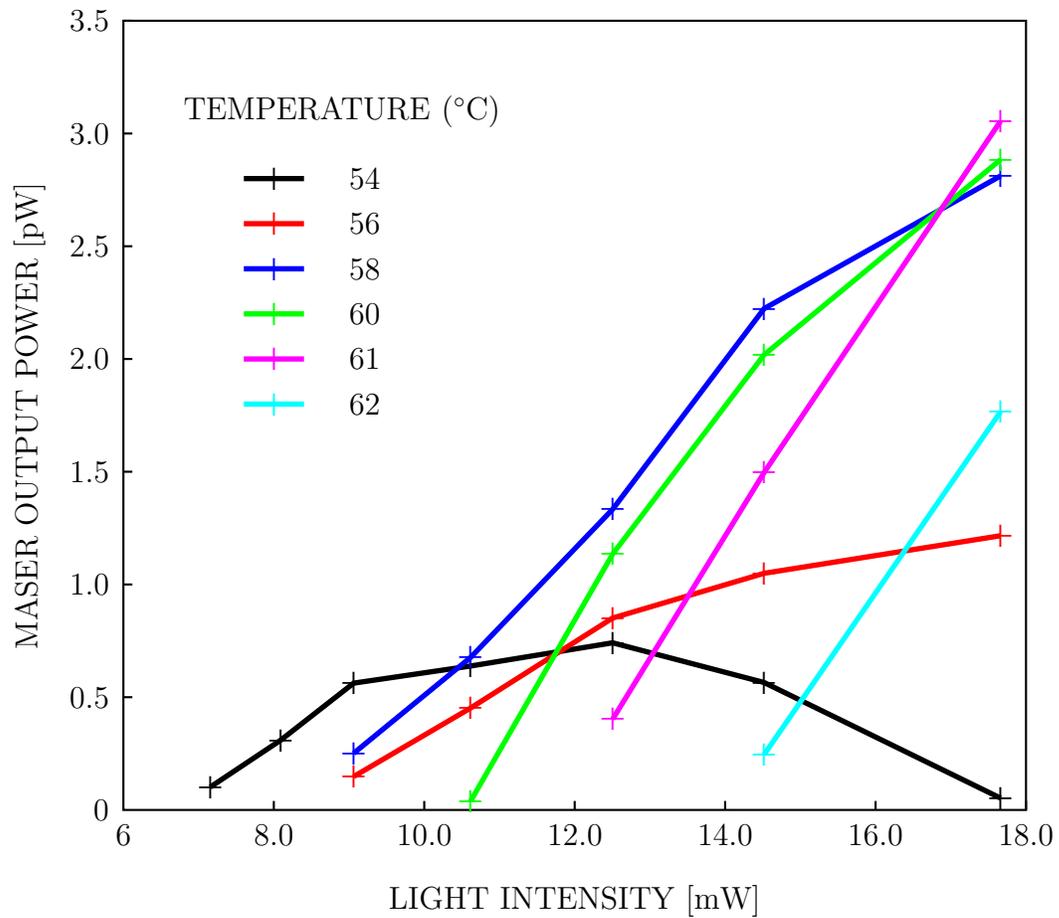
\section{Maser Frequency}

Modulating the laser frequency also results in a frequency modulation
of the maser signal.
Fig.~\ref{fig:sbvsf} gives the relative level of the modulation sidebands
when the
modulation frequency is changed. We see that they decrease as the
frequency modulation is increased at a rate of about 20 dB per decade.
Their effect can be minimized by a proper choice of the modulation frequency.
As an example for a modulation frequency of 20 kHz the curve shows
sidebands levels of about $-50$ dBc. This level could deteriorate the
short term stability to a level of about $3 \times 10^{-13} \tau^{-1}$
\cite{RUT78},
which is the same as the
stability
of the maser itself. Moreover, the maser signal is used to phase-lock
a quartz oscillator through a coherent receiver. It has been shown that
the optimum bandwidth of the phase-locked loop is around 1 kHz \cite{MIC86},
so the remaining spurious sidebands will be
filtered out \cite{VAN82}.

\begin{figure}
\psset{yunit=0.22cm,xunit=4cm}
\begin{pspicture}(-0.4,-60)(3.5,-10)
\psaxes[Dx=1.0,Oy=-55,Dy=5,axesstyle=frame,
      tickstyle=top,labels=y,ticks=xy](0,-55)(0,-55)(3,-10)
\rput(0,-57){0.1}
\rput(1,-57){1}
\rput(2,-57){10}
\rput(3,-57){100}
\rput(1.5,-60){MODULATION FREQUENCY [kHz]}
\rput*[c]{L}(-0.38,-33){SIDEBANDS LEVEL [dBc]}
\psline[linewidth=0.1pt,linecolor=green](0.30103,-55)(0.30103,-10)
\psline[linewidth=0.1pt,linecolor=green](0.477121254,-55)(0.477121254,-10)
\psline[linewidth=0.1pt,linecolor=green](0.60206,-55)(0.60206,-10)
\psline[linewidth=0.1pt,linecolor=green](0.69897,-55)(0.69897,-10)
\psline[linewidth=0.1pt,linecolor=green](0.77815,-55)(0.77815,-10)
\psline[linewidth=0.1pt,linecolor=green](0.8451,-55)(0.8451,-10)
\psline[linewidth=0.1pt,linecolor=green](0.90309,-55)(0.90309,-10)
\psline[linewidth=0.1pt,linecolor=green](0.954242,-55)(0.954242,-10)
\psline[linewidth=0.1pt,linecolor=green](1,-55)(1,-10)
\psline[linewidth=0.1pt,linecolor=green](1.30103,-55)(1.30103,-10)
\psline[linewidth=0.1pt,linecolor=green](1.477121254,-55)(1.477121254,-10)
\psline[linewidth=0.1pt,linecolor=green](1.60206,-55)(1.60206,-10)
\psline[linewidth=0.1pt,linecolor=green](1.69897,-55)(1.69897,-10)
\psline[linewidth=0.1pt,linecolor=green](1.77815,-55)(1.77815,-10)
\psline[linewidth=0.1pt,linecolor=green](1.8451,-55)(1.8451,-10)
\psline[linewidth=0.1pt,linecolor=green](1.90309,-55)(1.90309,-10)
\psline[linewidth=0.1pt,linecolor=green](1.954242,-55)(1.954242,-10)
\psline[linewidth=0.1pt,linecolor=green](2,-55)(2,-10)
\psline[linewidth=0.1pt,linecolor=green](2.30103,-55)(2.30103,-10)
\psline[linewidth=0.1pt,linecolor=green](2.477121254,-55)(2.477121254,-10)
\psline[linewidth=0.1pt,linecolor=green](2.60206,-55)(2.60206,-10)
\psline[linewidth=0.1pt,linecolor=green](2.69897,-55)(2.69897,-10)
\psline[linewidth=0.1pt,linecolor=green](2.77815,-55)(2.77815,-10)
\psline[linewidth=0.1pt,linecolor=green](2.8451,-55)(2.8451,-10)
\psline[linewidth=0.1pt,linecolor=green](2.90309,-55)(2.90309,-10)
\psline[linewidth=0.1pt,linecolor=green](2.954242,-55)(2.954242,-10)
\savedata{\mydata}[{
{0.30103,-14.16},
{0.47712,-17.67},
{0.69897,-21.84},
{0.8451,-24.67},
{1.0,-27},
{1.30103,-33.17},
{1.47712,-36},
{1.69897,-39.8},
{1.8451,-43.17},
{2,-46.17},
{2.30103,-50.3}}]
\dataplot[plotstyle=line,linewidth=2pt,linecolor=red]{\mydata}
\savedata{\mydata}[{
{0.30103,-14.16},
{0.47712,-17.67},
{0.69897,-21.84},
{0.8451,-24.67},
{1.0,-27},
{1.30103,-33.17},
{1.47712,-36},
{1.69897,-39.8},
{1.8451,-43.17},
{2,-46.17},
{2.30103,-50.3}}]
\dataplot[plotstyle=dots,showpoints=true,
     dotstyle=+,dotscale=2,linecolor=red]{\mydata}
\savedata{\mydata}[{{2.1,-19}}]
\dataplot[plotstyle=dots,showpoints=true,
     dotstyle=+,dotscale=2,linecolor=red]{\mydata}
\rput*[l](2.2,-19){SQUARE}
\savedata{\mydata}[{
{0.47712,-17},
{0.69897,-21.34},
{0.84509,-24.67},
{1.0,    -27.5},
{1.30103,-33},
{1.47712,-36.16},
{1.69897,-41.33},
{1.8451, -42.8},
{2.0,    -45.2}}]
\dataplot[plotstyle=line,linewidth=2pt,linecolor=blue]{\mydata}
\savedata{\mydata}[{
{0.47712,-17},
{0.69897,-21.34},
{0.84509,-24.67},
{1.0,    -27.5},
{1.30103,-33},
{1.47712,-36.16},
{1.69897,-41.33},
{1.8451, -42.8},
{2.0,    -45.2}}]
\dataplot[plotstyle=dots,showpoints=true,
   dotstyle=+,dotscale=2,linecolor=blue]{\mydata}
\savedata{\mydata}[{{2.1,-22}}]
\dataplot[plotstyle=dots,showpoints=true,
   dotstyle=+,dotscale=2,linecolor=blue]{\mydata}
\rput*[l](2.2,-22){SINE}
\savedata{\mydata}[{
{0.69897,-22.5},
{0.8451,-25},
{1.0,-28},
{1.30103,-33},
{1.47712,-37},
{1.69897,-41.2},
{1.8451,-43.37},
{2.0,-45.67},
{2.30103,-49.7}}]
\dataplot[plotstyle=line,linewidth=2pt,linecolor=green]{\mydata}
\savedata{\mydata}[{
{0.69897,-22.5},
{0.8451,-25},
{1.0,-28},
{1.30103,-33},
{1.47712,-37},
{1.69897,-41.2},
{1.8451,-43.37},
{2.0,-45.67},
{2.30103,-49.7}}]
\dataplot[plotstyle=dots,showpoints=true,dotstyle=+,dotscale=2,linecolor=green]{\mydata}
\savedata{\mydata}[{{2.1,-25}}]
\dataplot[plotstyle=dots,showpoints=true,dotstyle=+,dotscale=2,linecolor=green]{\mydata}
\rput*[l](2.2,-25){TRIANGLE}
\end{pspicture}
\caption{Sidebands level as a function of the modulation frequency for different modulation waveform.}
\label{fig:sbvsf}
\end{figure}
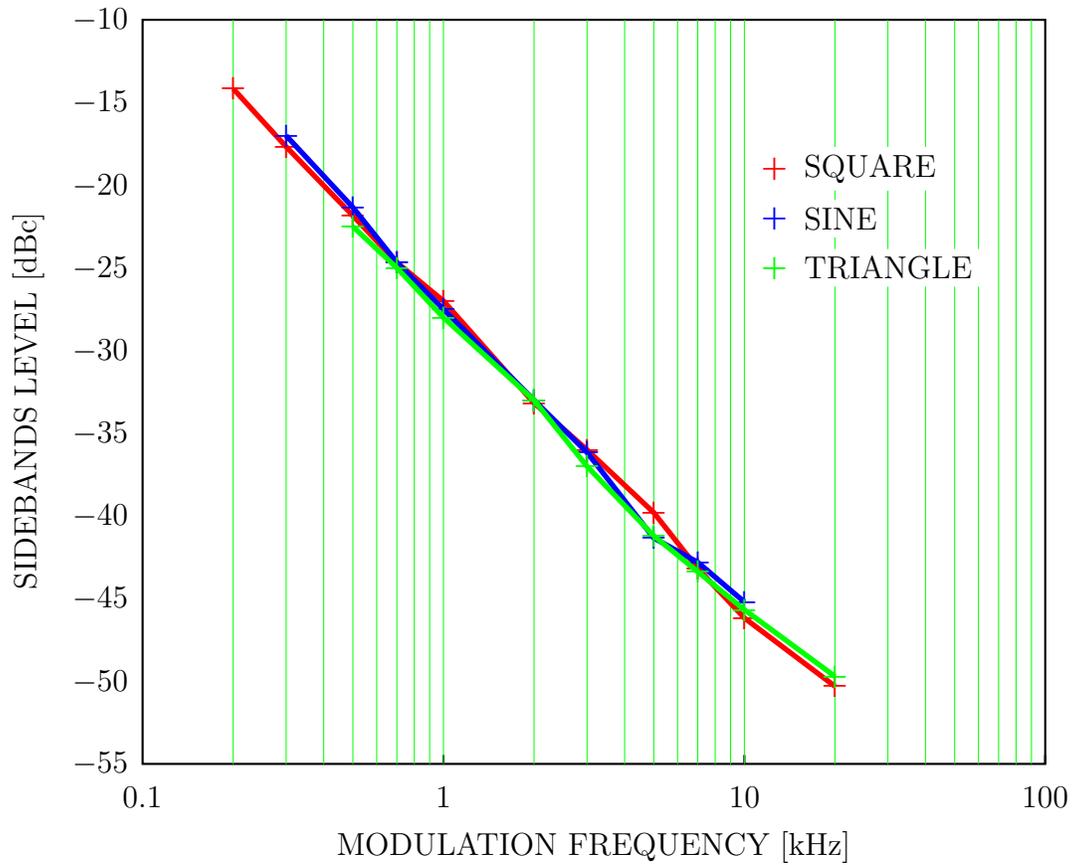

Fig.~\ref{fig:freq} shows the maser frequency when the laser is detuned.
Again the zero offset for the noise modulation
is arbitrary. We observe that the frequency dependence
of the curve is about the same for
all types of modulation. The measured pulling coefficient
$y_{m}/y_{l}$ is about $4.6 \times 10^{-3}$.
The frequency stability of the maser has been measured with the
laser unlocked, and showed a flicker floor level of
about $1 \times 10^{-11}$,
leading to a laser stability of $ 5 \times 10^{-9}$, which is typical for the
setup we used \cite{TET89}.

\begin{figure}
\psset{yunit=0.2cm,xunit=14cm}
\begin{pspicture}(-0.50,345)(0.41,400)
\psaxes[Ox=-0.4,Dx=0.1,Oy=350,Dy=5,axesstyle=frame,
      tickstyle=top,labels=all,ticks=all](-0.4,350)(-0.4,350)(0.4,400)
\rput(0,345){LASER DETUNING [GHz]}
\rput(-0.23,397){+ 6 834 688 000 [Hz]}
%
\rput*[c]{L}(-0.49,375){MASER FREQUENCY [Hz]}
\savedata{\mydata}[{
{-3.038E-1,3.5628654E+2},
{-2.921E-1,3.5648185E+2},
{-2.804E-1,3.5693031E+2},
{-2.687E-1,3.5739100E+2},
{-2.571E-1,3.5779946E+2},
{-2.454E-1,3.5791896E+2},
{-2.337E-1,3.5869769E+2},
{-2.220E-1,3.5910270E+2},
{-2.103E-1,3.5944723E+2},
{-1.986E-1,3.5996449E+2},
{-1.869E-1,3.6061357E+2},
{-1.753E-1,3.6140212E+2},
{-1.636E-1,3.6203646E+2},
{-1.519E-1,3.6274799E+2},
{-1.402E-1,3.6345129E+2},
{-1.285E-1,3.6471305E+2},
{-1.168E-1,3.6549990E+2},
{-1.051E-1,3.6589866E+2},
{-9.350E-2,3.6601574E+2},
{-8.180E-2,3.6654902E+2},
{-7.010E-2,3.6722974E+2},
{-5.840E-2,3.6861446E+2},
{-4.670E-2,3.6945861E+2},
{-3.500E-2,3.7051962E+2},
{-2.340E-2,3.7131754E+2},
{-1.170E-2,3.7172872E+2},
{0.000E+0,3.73029540E+2},
{1.170E-2,3.7411325E+2},
{2.340E-2,3.7579640E+2},
{3.510E-2,3.7714520E+2},
{4.680E-2,3.7876914E+2},
{5.840E-2,3.8045211E+2},
{7.010E-2,3.8163090E+2},
{8.180E-2,3.8211483E+2},
{9.350E-2,3.8258988E+2},
{1.052E-1,3.8378463E+2},
{1.169E-1,3.8482033E+2},
{1.286E-1,3.8583764E+2},
{1.402E-1,3.8691546E+2},
{1.519E-1,3.8823366E+2},
{1.636E-1,3.8913866E+2},
{1.753E-1,3.9034871E+2},
{1.870E-1,3.9175387E+2},
{1.987E-1,3.9240858E+2},
{2.104E-1,3.9274869E+2},
{2.220E-1,3.9379130E+2},
{2.337E-1,3.9435302E+2},
{2.454E-1,3.9505440E+2},
{2.571E-1,3.9607904E+2},
{2.688E-1,3.9676043E+2},
{2.805E-1,3.9776344E+2},
{2.921E-1,3.9859050E+2},
{3.038E-1,3.9927528E+2},
}]
\dataplot[plotstyle=dots,showpoints=true,
     dotstyle=+,dotscale=1.6,linecolor=red]{\mydata}
\savedata{\mydata}[{{0.15,370}}]
\dataplot[plotstyle=dots,showpoints=true,
     dotstyle=+,dotscale=1.6,linecolor=red]{\mydata}
\rput*[l](0.2,370){SQUARE}
\savedata{\mydata}[{
{-3.272E-1,3.5355430E+2},
{-3.155E-1,3.5418392E+2},
{-3.038E-1,3.5478113E+2},
{-2.921E-1,3.5578441E+2},
{-2.804E-1,3.5651701E+2},
{-2.688E-1,3.5672445E+2},
{-2.571E-1,3.5712620E+2},
{-2.454E-1,3.5819921E+2},
{-2.337E-1,3.5916165E+2},
{-2.220E-1,3.5986011E+2},
{-2.103E-1,3.6027966E+2},
{-1.986E-1,3.6067144E+2},
{-1.870E-1,3.6135889E+2},
{-1.753E-1,3.6194753E+2},
{-1.636E-1,3.6232514E+2},
{-1.519E-1,3.6288587E+2},
{-1.402E-1,3.6361891E+2},
{-1.285E-1,3.6455163E+2},
{-1.168E-1,3.6477822E+2},
{-1.052E-1,3.6530939E+2},
{-9.350E-2,3.6564611E+2},
{-8.180E-2,3.6617062E+2},
{-7.010E-2,3.6648803E+2},
{-5.840E-2,3.6784187E+2},
{-4.670E-2,3.6863238E+2},
{-3.510E-2,3.6982338E+2},
{-2.340E-2,3.6977938E+2},
{-1.170E-2,3.7026672E+2},
{0.000E+0,3.7081326E+2},
{1.170E-2,3.7154250E+2},
{2.340E-2,3.7247307E+2},
{3.510E-2,3.7397657E+2},
{4.670E-2,3.7501730E+2},
{5.840E-2,3.7522326E+2},
{7.010E-2,3.7640621E+2},
{8.180E-2,3.7777216E+2},
{9.350E-2,3.7834712E+2},
{1.052E-1,3.7910791E+2},
{1.169E-1,3.8013355E+2},
{1.285E-1,3.8118700E+2},
{1.402E-1,3.8282386E+2},
{1.519E-1,3.8382814E+2},
{1.636E-1,3.8441098E+2},
{1.753E-1,3.8573675E+2},
{1.870E-1,3.8636186E+2},
{1.987E-1,3.8706516E+2},
{2.103E-1,3.8817166E+2},
{2.220E-1,3.8968355E+2},
{2.337E-1,3.9075641E+2},
{2.454E-1,3.9201606E+2},
{2.571E-1,3.9316251E+2},
{2.688E-1,3.9379529E+2},
{2.804E-1,3.9453311E+2},
{2.921E-1,3.9511377E+2},
{3.038E-1,3.9580719E+2},
{3.155E-1,3.9722578E+2},}]
\dataplot[plotstyle=dots,showpoints=true,
   dotstyle=+,dotscale=1.6,linecolor=blue]{\mydata}
\savedata{\mydata}[{{0.15,367}}]
\dataplot[plotstyle=dots,showpoints=true,
   dotstyle=+,dotscale=1.6,linecolor=blue]{\mydata}
\rput*[l](0.2,367){SINE}
\savedata{\mydata}[{
{-2.688E-1,3.5104131E+2}
{-2.571E-1,3.5158406E+2}
{-2.454E-1,3.5186796E+2}
{-2.337E-1,3.5204319E+2}
{-2.220E-1,3.5302138E+2}
{-2.104E-1,3.5385165E+2}
{-1.987E-1,3.5450953E+2}
{-1.870E-1,3.5543738E+2}
{-1.753E-1,3.5628378E+2}
{-1.636E-1,3.5687778E+2}
{-1.519E-1,3.5752561E+2}
{-1.403E-1,3.5843244E+2}
{-1.286E-1,3.5916829E+2}
{-1.169E-1,3.6003702E+2}
{-1.052E-1,3.6111772E+2}
{-9.350E-2,3.6241206E+2}
{-8.180E-2,3.6381187E+2}
{-7.010E-2,3.6489196E+2}
{-5.850E-2,3.6574325E+2}
{-4.680E-2,3.6665820E+2}
{-3.510E-2,3.6755066E+2}
{-2.340E-2,3.6874166E+2}
{-1.170E-2,3.6970196E+2}
{0.000E+0,3.7092514E+2}
{1.170E-2,3.7200648E+2}
{2.330E-2,3.7289084E+2}
{3.500E-2,3.7390945E+2}
{4.670E-2,3.7474963E+2}
{5.840E-2,3.7560134E+2}
{7.010E-2,3.7586263E+2}
{8.180E-2,3.7710596E+2}
{9.340E-2,3.7818042E+2}
{1.051E-1,3.7868498E+2}
{1.168E-1,3.7925480E+2}
{1.285E-1,3.7963092E+2}
{1.402E-1,3.8073103E+2}
{1.519E-1,3.8197006E+2}
{1.636E-1,3.8346124E+2}
{1.752E-1,3.8439241E+2}
{1.869E-1,3.8509313E+2}
{1.986E-1,3.8665745E+2}
{2.103E-1,3.8802771E+2}
{2.220E-1,3.8884138E+2}
{2.337E-1,3.8983796E+2}
{2.454E-1,3.9087019E+2}
{2.570E-1,3.9252123E+2}
{2.687E-1,3.9335589E+2}
}]
\dataplot[plotstyle=dots,showpoints=true,
     dotstyle=+,dotscale=1.6,linecolor=green]{\mydata}
\savedata{\mydata}[{{0.15,364}}]
\dataplot[plotstyle=dots,showpoints=true,
     dotstyle=+,dotscale=1.6,linecolor=green]{\mydata}
\rput*[l](0.2,364){TRIANGLE}
\savedata{\mydata}[{
{-3.155E-1,3.5033424E+2}
{-3.038E-1,3.5111927E+2}
{-2.921E-1,3.5187902E+2}
{-2.804E-1,3.5238960E+2}
{-2.688E-1,3.5341861E+2}
{-2.571E-1,3.5438056E+2}
{-2.454E-1,3.5555116E+2}
{-2.337E-1,3.5653009E+2}
{-2.220E-1,3.5733314E+2}
{-2.103E-1,3.5797603E+2}
{-1.986E-1,3.5912897E+2}
{-1.870E-1,3.5977170E+2}
{-1.753E-1,3.6058031E+2}
{-1.636E-1,3.6154534E+2}
{-1.519E-1,3.6270465E+2}
{-1.402E-1,3.6340189E+2}
{-1.285E-1,3.6423857E+2}
{-1.169E-1,3.6518118E+2}
{-1.052E-1,3.6604138E+2}
{-9.350E-2,3.6712327E+2}
{-8.180E-2,3.6753943E+2}
{-7.010E-2,3.6846488E+2}
{-5.840E-2,3.6982234E+2}
{-4.670E-2,3.7064518E+2}
{-3.510E-2,3.7190044E+2}
{-2.340E-2,3.7291631E+2}
{-1.170E-2,3.7365371E+2}
{0.000E+0,3.7469206E+2}
{1.170E-2,3.7548634E+2}
{2.340E-2,3.7624109E+2}
{3.510E-2,3.7707161E+2}
{4.670E-2,3.7788968E+2}
{5.840E-2,3.7972641E+2}
{7.010E-2,3.8123202E+2}
{8.180E-2,3.8160432E+2}
{9.350E-2,3.8270938E+2}
{1.052E-1,3.8343446E+2}
{1.169E-1,3.8426504E+2}
{1.285E-1,3.8534771E+2}
{1.402E-1,3.8622901E+2}
{1.519E-1,3.8689936E+2}
}]
\dataplot[plotstyle=dots,showpoints=true,
     dotstyle=+,dotscale=1.6,linecolor=black]{\mydata}
\savedata{\mydata}[{{0.15,361}}]
\dataplot[plotstyle=dots,showpoints=true,
     dotstyle=+,dotscale=1.6,linecolor=black,fillcolor=black]{\mydata}
\rput*[l](0.2,361){NOISE}
\end{pspicture}
\caption{Maser frequency as a function of the average laser detuning.}
\label{fig:freq}
\end{figure}
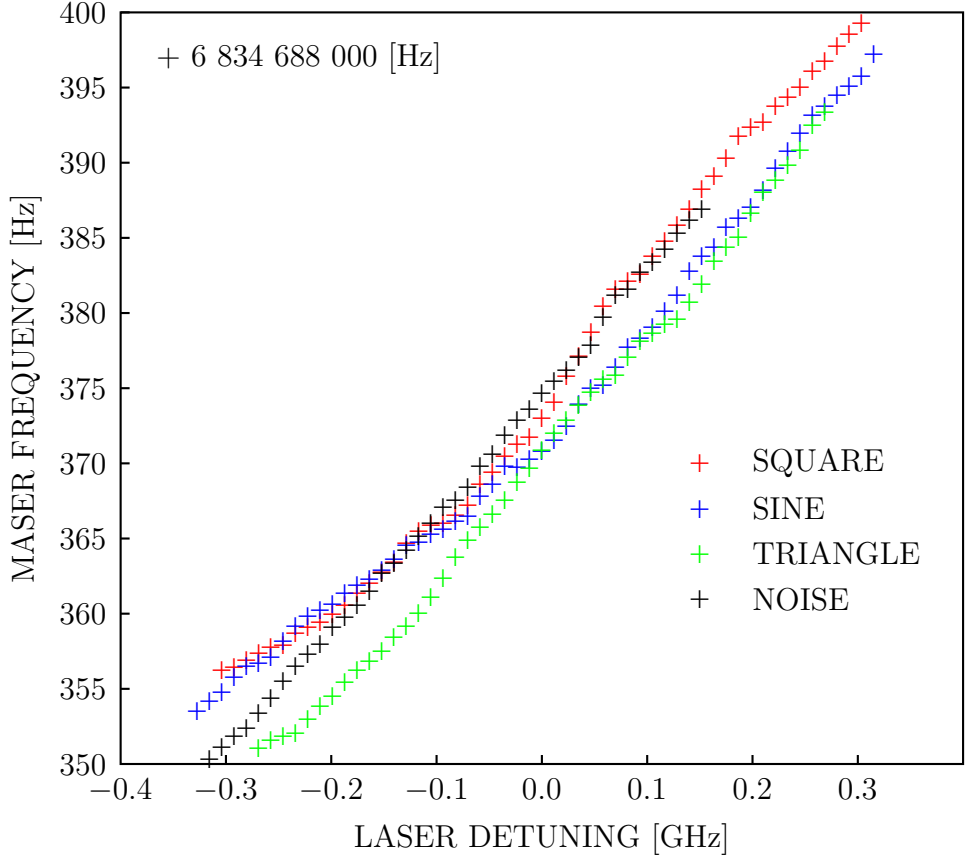

This stresses out the importance of the laser frequency stabilization
in order to achieve good performances.
The frequency locking of the
laser improves its stability by two
orders of magnitude \cite{TET89}, which would bring the maser
stability to about $5 \times 10^{-13}$.
Recently, techniques have been reported to further increase
the stability of the laser \cite{LAB86}.
These techniques along with a cavity tuning
system based on light intensity modulation could result in presently
unachievable performance.

\section{Conclusion}

Laser diode used as a pumping source on the rubidium maser will
permit the construction of a more compact frequency standard.
The design of the maser involves the
broadening of the laser spectrum in order to maximize the maser power.
Periodic deterministic modulations are preferable to random modulations,
since they produce higher maser power and
permit the frequency locking technique
on an external absorption cell. It has been
shown that the modulation frequency has to be high enough in order
to ensure high maser power and low frequency modulation spurious.
Measurements have shown a relative laser
frequency pulling over the relative maser frequency of about $4.6 \times
10^{-3}$.

\section*{Acknowledgements}
The authors wish to thank Prof.
Yang Shi-Qi on leave from South China Normal University,
and S. Th\'{e}riault, M. Chamberland and M. Levesque from Laval University,
for their help and discussions.
This research was supported by Natural Sciences and Engineering
Research Council, Canada, and Fonds F.C.A.R., Qu\'{e}bec.


%

\begin{thebibliography}{9}

\bibitem{DEC84} \'{E}. de Clerq {\em et al.},
 ``Laser diode optically pumped caesium beam,''
        {\em Journal de physique}, Vol. 45, pp. 239--247, Feb. 1984.

\bibitem{OHS88} S.I. Ohshima {\em et al.} ``Development of an Optically
                Pumped Cs Frequency Standard at the NRLM,''
                {\em IEEE Trans. on Instrumentation and Measurement},
                Vol. 37, pp. 409--13, Sept. 1988.

\bibitem{TET85} M. T\^{e}tu {\em et al.},
         ``Experimental Results On a Frequency
           Standard Based on a Rubidium 87 Maser,''
           {\em Proc. 36th Annual Symposium on Frequency Control},
           Philadelphia, U.S.A., pp. 64--71, May 1985.

\bibitem{TET89} M. T\^{e}tu {\em et al.}, ``Multiwavelength Sources
           Using Laser Diodes Frequency-Locked to Atomic Resonances''
               {\em Journal of Lightwave Technology}, Vol. 7,
               pp. 1540-8, Oct. 1989, DOI:10.1109/50.39095.
Available: [online] .

\bibitem{RUT78} J. Rutman, ``Characterization of Phase and Frequency
                             Instabilities in Precision Frequency Sources:
                             Fifteen Years of Progress''
               {\em Proceedings of IEEE}, Vol. 66,
               pp. 1048-75, Sept. 1978,
Available: [online] .

\bibitem{MIC86} A. Michaud, ``Etude de la stabilit\'{e} de la fr\'{e}quence
         d'un \'{e}talon de fr\'{e}quence bas\'{e} sur un maser \`{a}
          rubidium 87:
         Influence du r\'{e}cepteur coh\'{e}rent.'',
         M. Sc. Thesis, Universit\'{e} Laval,Qu\'{e}bec,Canada, May 1986,
	 Available: [online] .

\bibitem{VAN82} J. Vanier and M. T\^{e}tu ``Phase-Locked Loops Used with
                Masers: Atomic Frequency Standards,''
                {\em IEEE Trans. on Communications},
                Vol. 30, pp. 2355--2361, Oct. 1982, Available: 
[online] .

\bibitem{LAB86} M. de Labachelerie and P. Cerez,
         `Frequency locking of a 850 nm
         external-cavity semiconductor laser on a Doppler-free Cs-D2 line'',
         Proc. SPIE, vol. 701, ECOOSA'86, Florence 1986.

\end{thebibliography}
\end{document}